\documentclass[twocolumn,trackchanges]{aastex7}
\usepackage{lmodern}
\usepackage{ae,aecompl,multirow}
\usepackage{threeparttable}
\usepackage{booktabs}
\usepackage{url}
\usepackage{amsmath}
\usepackage{graphics,comment}
\usepackage{textcomp} % For \textendash support
\usepackage{hyperref} % Load hyperref for cross-referencing

\usepackage{booktabs}  % For better table formatting
\DeclareUnicodeCharacter{2212}{\textminus}

\begin{document}

\title{WISE-VAULT: Decade-long Mid-Infrared Variability of WISE AGNs}
\author[0000-0002-8377-9667]{Suvendu Rakshit}
%\email[show]{suvenduat@gmail.com}
\ifdefined\HCode
  \email{suvenduat@gmail.com}
\else
  \email[show]{suvenduat@gmail.com}
\fi
\affiliation{Aryabhatta Research Institute of Observational Sciences, Nainital\textendash263001, Uttarakhand, India}

\begin{abstract}
Mid-infrared (MIR) variability provides a powerful probe of accretion processes and dust reprocessing in active galactic nuclei (AGNs), as it is less affected by extinction than optical flux. Using multi-epoch photometry from \textit{WISE} and NEOWISE spanning over a decade, we investigate the MIR variability of $\sim$4.5 million color-selected AGNs. Variability is quantified using maximum--minimum amplitudes ($\Delta W1$), excess variance, reduced $\chi^2$, and structure function (SF) analysis over timescales of 0.5--10 years. We identify 632,939 variable AGNs, including 54,897 sources that exhibit large-amplitude variability ($\Delta W1 > 0.5$ mag) and 624 with extreme intrinsic variability ($A_{\rm var} > 0.5$ mag). A subset of these sources shows flare-like behavior, identifying them as candidate changing-look AGNs or tidal disruption events that warrant further spectroscopic and multi-wavelength follow-up. The SF analysis shows radio-detected AGNs exhibit the highest variability amplitudes, followed by X-ray detected and the overall AGN sample. The variability amplitude is strongly anti-correlated with the Eddington ratio, indicating enhanced variability in lower accretion rate systems. A weak positive dependence on radio-loudness further suggests a contribution from non-thermal emission in radio sources. These results demonstrate that MIR variability provides a robust diagnostic of accretion physics and an efficient means of identifying extreme and rapidly evolving AGN phenomena. The full catalog of variable AGNs is made publicly available.
\end{abstract}

% Select between one and six entries from the list of approved keywords.
% Don't make up new ones.


\section{Introduction}

Accretion of matter onto a supermassive black hole is widely accepted as the primary mechanism powering active galactic nuclei (AGNs), producing luminous emission across the electromagnetic spectrum and giving rise to pronounced variability on timescales ranging from hours to decades \citep{1969Natur.223..690L,1973A&A....24..337S}. This variability is a fundamental consequence of the dynamical nature of the accretion flow and its interaction with surrounding structures, including the X-ray–emitting corona and circumnuclear material. Because the characteristic spatial scales of the accretion disk and reprocessing regions are far below the resolution limits of direct imaging, temporal variability serves as a powerful indirect probe of the physical conditions, geometry, and size scales of AGN central engines \citep[e.g.,][]{1963ApJ...138...30M,1997ARA&A..35..445U}.

The physical origin of AGN variability is commonly attributed to stochastic processes within the accretion disk, including fluctuations in the mass accretion rate, magneto-rotational turbulence, thermal instabilities, and variable irradiation by a compact X-ray–emitting corona \citep{ShakuraSunyaev1976,1993ApJ...412..254C,2011ApJ...727L..24D,2014ApJ...783..105R}. These processes produce correlated variability over a wide range of timescales and motivate the use of statistical descriptions such as damped random walk (DRW) and related stochastic models. Such models have been successfully applied to optical AGN light curves from large surveys and monitoring campaigns, revealing empirical correlations between variability properties and fundamental AGN parameters, including black hole mass, accretion rate, and luminosity \citep[see][]{2009ApJ...698..895K,2010ApJ...716..530K,2010ApJ...721.1014M,2016ApJ...826..118K,2017ApJ...842...96R}.

While optical variability directly traces emission from the accretion disk, infrared variability probes a physically distinct and spatially extended component of the AGN structure. In the MIR, the emission is dominated by thermal radiation from dust in the circumnuclear torus that absorbs ultraviolet and optical photons and re-emits them at longer wavelengths \citep{1993ARA&A..31..473A,2015ARA&A..53..365N}. Variability in the accretion disk luminosity is therefore expected to drive infrared variability through dust reprocessing, with characteristic time delays and smoothing determined by the light-travel time across the torus and the thermal response of the dust grains \citep{1987ApJ...320..537B,2014ApJ...788..159K}. Consequently, the amplitude and timescale dependence of infrared variability provide direct constraints on the geometry, size, and radial distribution of the dusty environment.

Early observational studies established that AGNs exhibit measurable variability at infrared wavelengths and that such variability is broadly correlated with optical emission, consistent with a reprocessing origin \citep[e.g.,][]{1992MNRAS.256P..23G,1999AJ....118...35N,2011ApJ...737..105K,2014ApJ...788..159K}. However, these studies were limited by small sample sizes, heterogeneous cadences, and relatively short temporal baselines, restricting their ability to probe long-term variability and population-level trends. The advent of wide-field, space-based infrared surveys has fundamentally transformed this field. Recent studies have demonstrated that infrared variability is common among AGNs and exhibits systematic dependencies on luminosity, wavelength, and rest-frame timescale \citep{2016ApJ...817..119K,2018ApJ...864...87S,2019ApJ...886...33L,2019MNRAS.483.2362R,2020ApJ...889...46S,2024ApJ...977..102P,2026ApJ..1000..245K}. Long-term WISE monitoring has also enabled detailed studies of MIR variability and dust reverberation in AGNs. \citet{2019ApJ...886...33L} developed methods to construct WISE MIR light curves and measure dust-reverberation lags in PG quasars, accounting for sparse sampling and photometric limitations. Subsequent studies combining long-term optical and WISE light curves measured MIR--optical lags and established the dust-lag--luminosity relation over a broad luminosity range \citep[e.g.,][]{2020ApJ...900...58Y,2024ApJ...968...59M,2025ApJ...993..203T}, further demonstrating the connection between optical variability and delayed MIR emission from circumnuclear dust. \citet{2023ApJ...950..122L} showed that ensemble optical and MIR structure functions can be used to characterize the smoothing and geometric dilution of variability by the dusty torus and to constrain its effective size.

Beyond characterizing typical variability, MIR time-domain studies have proven particularly powerful for identifying extreme and rapidly evolving nuclear phenomena. Changing-look AGNs (CL-AGNs), which undergo dramatic transitions in their optical spectral type on timescales of years, are increasingly interpreted as manifestations of large-amplitude changes in accretion rate rather than obscuration alone \citep[e.g.,][]{2015ApJ...800..144L,2016ApJ...826..188R,2019ApJ...874....8M}. In such events, the MIR emission often responds with delayed but significant brightening or fading, reflecting changes in dust heating and, in some cases, dust destruction or reformation \citep{2017ApJ...846L...7S,2018ApJ...862..109Y}. Similarly, tidal disruption events (TDEs), in which a star is tidally disrupted by a dormant or weakly active supermassive black hole, can produce luminous infrared echoes as the intense flare irradiates surrounding dust \citep[e.g.,][]{2016ApJ...832..188D,2016ApJ...829...19V,2018Sci...361..482M,2021ApJS..252...32J}. In both cases, MIR variability provides a unique and often less obscured diagnostic compared to optical emission, which can be significantly affected by extinction and host-galaxy contamination. Moreover, using mid-IR data, \citet{2026ApJ..1000..245K} demonstrate that AGN candidates can be identified even in optically inactive hosts, with a low contamination rate ($\sim$1\%) from transients such as TDEs and supernovae.

Despite these advances, a comprehensive characterization of long-term MIR variability across large AGN samples remains incomplete. In particular, uniform statistical analyses that explicitly quantify variability as a function of timescale—such as structure function measurements—are still relatively scarce in the infrared regime. Such analyses are essential for distinguishing between stochastic accretion-driven variability, dust reprocessing effects, and rare transient phenomena, and for enabling direct comparisons with variability properties observed at shorter wavelengths.

In this paper, we present a comprehensive study of long-term MIR variability in a large sample of AGNs selected from the MIR color-selected AGN catalog of \citet{2018ApJS..234...23A} using their R90 reliability criterion, corresponding to a 90\% reliability threshold for AGN selection based on WISE colors. Utilizing multi-epoch photometry from \textit{WISE} and NEOWISE, we perform a suite of variability analyses, including statistical variability diagnostics and ensemble structure function measurements, to characterize the amplitude, timescale dependence, and population-level behavior of infrared variability. In addition, our analysis enables the identification of a significant population of highly variable sources, which are strong candidates for extreme accretion-driven events such as CL-AGN and TDEs. We compile the key variability properties derived in this work into the \textit{WISE Variability of Active Galaxies over Long Timescales} (``WISE-VAULT'') catalog, which provides a valuable resource for studying AGN variability and identifying extreme accretion events. The paper is organized as follows: Section~\ref{sec:data} describes the data and sample selection, Section~\ref{sec:analysis} outlines the variability analysis methods, Section~\ref{sec:result} presents the main results, and Section~\ref{sec:discussion} discusses their implications with a conclusion in Section~\ref{sec:conclusion}.

\section{Sample and Data} \label{sec:data}

\begin{figure}
    \centering
    \includegraphics[width=9cm, height=7cm]{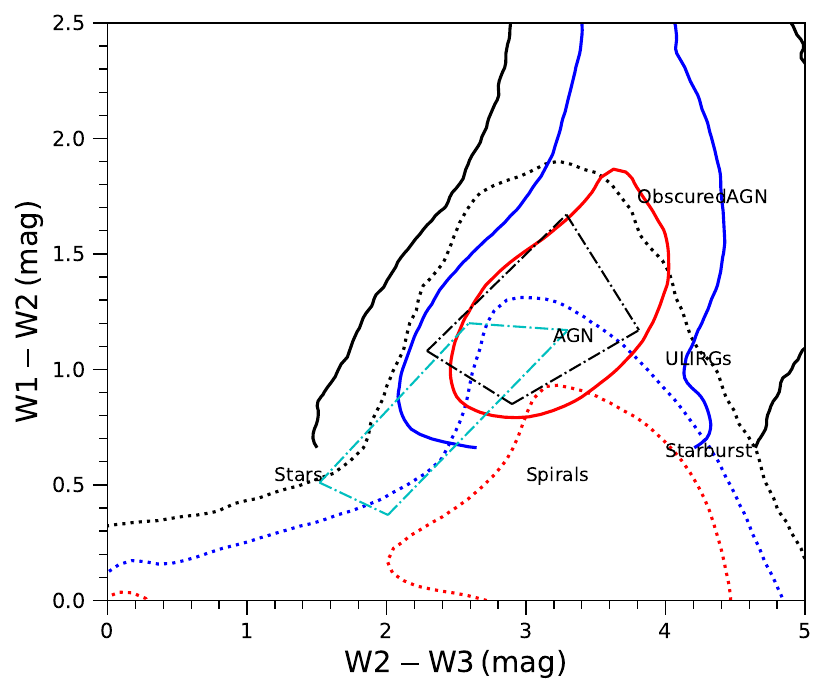}
    \caption{WISE color--color diagram for the AGN sample constructed from the WISE Single Exposure Source Table. The dotted contours indicate the $1\sigma$ (red), $2\sigma$ (blue), and $3\sigma$ (black) levels for different classes of astronomical objects; their locations are marked in the diagram. For reference, the distribution of color-selected AGNs from the R90 catalog of \citet{2018ApJS..234...23A} is represented by solid contours.}
    \label{fig:wise_color}
\end{figure}

Our AGN sample is drawn from the \textit{Wide-field Infrared Survey Explorer} \citep[WISE;][]{2010AJ....140.1868W} MIR color-selected AGN catalog constructed by \citet{2018ApJS..234...23A} using the R90 reliability criterion. The R90 selection is designed to provide a 90\% reliability AGN sample based on W1 (3.4~$\mu$m) and W2 (4.6~$\mu$m) colors, effectively separating luminous AGNs from stars and normal galaxies over a wide redshift range. This reliability was established through a detailed analysis of UV-to-near-IR spectral energy distributions of sources in the 9~deg$^2$ Boötes field. The resulting R90 catalog contains 4,543,530 AGN candidates over the extragalactic sky, with an AGN candidate surface density of $130\pm4$~deg$^{-2}$, providing one of the largest uniformly selected MIR AGN samples currently available. This selection is particularly well-suited for infrared variability studies, as it identifies sources whose MIR emission is dominated by hot dust heated by the central engine and is therefore sensitive to variability driven by changes in accretion activity. In Figure~\ref{fig:wise_color},  we show the distribution of the parent AGN sample in the WISE color--color diagram, together with the color distributions of other astronomical source populations for comparison. The figure illustrates the location of the parent sample in WISE color space relative to these populations, while the solid contours show the distribution of color-selected AGNs from the R90 catalog of \citet{2018ApJS..234...23A}.

We cross-matched the parent R90 AGN catalog with the WISE database using a matching radius of 3\arcsec. This radius is comparable to the WISE point-spread function (PSF), whose full width at half maximum (FWHM) is approximately 6.1\arcsec in W1 and 6.4\arcsec in W2 \citep{2010AJ....140.1868W}. The adopted radius therefore provides a balance between recovering genuine optical counterparts and minimizing chance associations. The WISE mission provides multi-epoch photometry from several operational phases, including the cryogenic mission (2010), the post-cryogenic phase, and the subsequent NEOWISE reactivation survey. After the primary mission concluded in 2011, WISE was reactivated as NEOWISE in December 2013 and has continued surveying the sky until August 2024. In this work, we combine photometric measurements from the AllWISE data release with the NEOWISE-R release (2023), which includes observations obtained between December 2013 and December 2022 in the W1 (3.4~$\mu$m) and W2 (4.6~$\mu$m) bands. Together, these data provide a temporal baseline exceeding a decade. Owing to WISE’s scanning strategy, most sky locations are observed in discrete six-month intervals (hereafter “epochs”), with individual exposures separated by approximately 1.5 hours within each epoch. By combining AllWISE and NEOWISE-R photometry, we construct decade-long light curves suitable for robust statistical analysis of MIR variability of AGNs.

Several factors can introduce spurious variability into WISE photometry, including scattered moonlight, contamination from nearby sources, image artifacts, and reduced image quality. To ensure robust variability measurements, we therefore restrict our analysis to high-quality single-exposure detections that are free from known contamination flags. Specifically, we apply the following criteria: \texttt{saa\_sep > 0} to exclude exposures affected by the South Atlantic Anomaly; \texttt{moon\_masked = 0000} to avoid moon contamination; \texttt{qi\_fact > 1} to retain frames with acceptable image quality; \texttt{nb < 3} and \texttt{na = 0} to remove blended or artifact-affected sources; \texttt{cc\_flags = 0000} to exclude known image artifacts; \texttt{w1rchi2\_ep < 3} to eliminate poor PSF fits; and \texttt{w1sigmpro\_ep < 0.2} mag to ensure reliable photometric uncertainties. We note that blending and host-galaxy emission may still affect the measured WISE fluxes. Unresolved nearby sources can dilute or add scatter to the measured variability, while host-galaxy emission can dilute the intrinsic AGN variability, particularly for lower-luminosity sources. These effects are not explicitly corrected for and may affect the measured variability amplitudes.

To construct long-term light curves, we group all qualified single-exposure measurements within each six-month observing interval into discrete epochs, requiring at least two measurements per epoch, and compute their weighted mean magnitudes. This epoch-averaging process reduces short-term stochastic noise and emphasizes variability on timescales of years. The uncertainty of the epoch-averaged magnitude is estimated following \citet{2019ApJ...886...33L}, combining the photometric errors and the observed scatter within each epoch as
 
\begin{equation}\label{eq:error}
\sigma_{\mathrm{int}}^{2}
=
\frac{1}{N-1}\sum_{j=1}^{N}
\left(f_j - \bar{f_{epoch}} \right)^{2}
\;+\;
\frac{1}{N^{2}}\sum_{j=1}^{N}
\sigma_{j,\mathrm{phot}}^{2}
\;+\;
\frac{1}{N}\,\sigma_{\mathrm{sys}}^{2},
\end{equation}
Here, $N$ is the number of data points within a 6-month window. $f_j$, $\sigma_{j,\mathrm{phot}}$, and $\bar{f_{epoch}}$ are the flux, its uncertainty for the $j$-th individual exposure, and the average flux within an epoch, respectively. $\sigma_{\mathrm{sys}}$ is the systematic uncertainty due to the instability of the instrument. \citet{2019ApJ...886...33L} estimated the $\sigma_{\mathrm{sys}}$ for ALLWISE and NEOWISE surveys using mid-IR calibration stars and found $\sigma_{\mathrm{sys}}$ to be $\sim0.029$ mag and $\sim0.016$ mag for ALLWISE and NEOWISE measurements. \citet{2023ApJ...958..135S} conducted an extensive investigation of the structure function using non-variable sources and found that the uncertainty estimated from Equation~\ref{eq:error} is overestimated. They showed that the true uncertainty is better represented by $\sigma_{\mathrm{true}} = \sigma_{\mathrm{int}}/\sqrt{N}$. 

To empirically assess the adopted epoch uncertainty, we constructed a control sample from the MPA-JHU DR8 catalog\footnote{\url{https://www.sdss4.org/dr16/spectro/galaxy_mpajhu/}} by selecting sources classified as \texttt{SPECTROTYPE=GALAXY} that are also present in our WISE sample. We applied the same WISE quality cuts and epoch construction procedure as for the AGN sample and calculated the ratio
\begin{equation}
R_{\rm err,true} = \frac{S^2}{\langle \sigma_{\rm true}^2 \rangle},
\end{equation}
where
\begin{equation}
S^2 = \frac{1}{N-1}\sum_{i=1}^{N}(m_i-\bar{m})^2,
\end{equation}
is the sample variance of the epoch magnitudes, with $N$ being the number of epochs, $m_i$ the magnitude measured in the $i$-th epoch, and $\bar{m}$ the mean magnitude. For a sample with correctly estimated uncertainties and negligible intrinsic variability, $R_{\rm err,true}$ is expected to be approximately unity. We find that $R_{\rm err,true}$ decreases systematically with increasing W1 magnitude, with values of 2.20--0.99 (median 1.5) across the magnitude bins (from brightest to faintest). This indicates that the corrected uncertainty ($\sigma_{\mathrm{true}}$) prescription provides a substantially better description of the observed scatter than the original $\sigma_{\rm int}$, although a residual dependence on source brightness remains, particularly for the brighter sources. For comparison, using the uncorrected $\sigma_{\rm int}$ gives $R_{\rm err}\simeq0.18$--0.12 (median 0.14) across the same magnitude range, indicating that $\sigma_{\rm int}$ substantially overestimates the observed scatter. Accordingly, we adopt $\sigma_{\mathrm{true}}$ as the uncertainty at each epoch in the final light curve. 

To ensure reliable characterization of long-term variability, we further restrict our analysis to sources with at least 16 well-sampled epochs. This requirement guarantees sufficient temporal coverage over the full WISE baseline and allows for robust statistical measurements of variability amplitude and structure function properties. After applying this criterion, the final sample used for subsequent variability analysis consists of 1,898,589 AGNs. 

\section{Variability Analysis}
\label{sec:analysis}

To characterize the MIR variability properties of our AGN sample, we adopt a two-step approach. First, we identify intrinsically variable sources using statistical variability diagnostics that account for photometric uncertainties. Second, we quantify the variability amplitude and its dependence on timescale using the structure function formalism.

\subsection{Detection of Intrinsic Variability}

\subsubsection{Chi-square Variability Test}

For each light curve, we evaluate deviations from a constant-flux hypothesis using the chi-square statistic \citep[e.g.,][]{1996ApJ...473..763M,2017ApJ...849..110S},

\begin{equation}
\chi^2 = \sum_{i=1}^{N}
\frac{\left(m_i - \langle m \rangle \right)^2}{\sigma_i^2},
\label{eq:chi2}
\end{equation}

where $m_i$ and $\sigma_i$ are the observed magnitude and its uncertainty at epoch $i$, $\langle m \rangle$ is the weighted mean magnitude, and $N$ is the total number of epochs in a light curve. Note that all variability analyses in this work are performed consistently in the magnitude scale.

Assuming no intrinsic variability, $\chi^2$ follows a chi-square distribution with $N-1$ degrees of freedom. Therefore, we compute the cumulative probability

\begin{equation}
P_{\rm var} = F_{\chi^2}(\chi^2; N-1),
\label{eq:Pvar}
\end{equation}

where $F_{\chi^2}$ is the cumulative chi-square distribution. We then define the variability parameter as

\begin{equation}
V_{\rm par} = -\log_{10}\left(1 - P_{\rm var}\right).
\label{eq:Vpar}
\end{equation}

Following \citet{2017ApJ...849..110S}, sources are classified as variable if $V_{\rm par} > 1.3$, corresponding to a $\sim95\%$ probability that the observed variability is inconsistent with a constant-flux model. Hence, a larger $V_{\rm par}$ indicates stronger variability. 

\subsubsection{Intrinsic Variability Amplitude}

To quantify the intrinsic variability amplitude, we adopt the excess variance formalism commonly used in AGN variability studies \citep[e.g.,][]{2003MNRAS.345.1271V,2017ApJ...842...96R}.

%For a light curve with $N$ epochs, the total variance is

%\begin{equation}
%S^2 = \frac{1}{N-1}
%\sum_{i=1}^{N} \left(m_i - \langle m \rangle \right)^2 .
%\end{equation}

The mean square photometric uncertainty is

\begin{equation}
\langle \sigma_i^2 \rangle =
\frac{1}{N} \sum_{i=1}^{N} \sigma_i^2 .
\label{eq:sigma}
\end{equation}

The intrinsic (noise-corrected) variance is then

\begin{equation}
\sigma_{\rm XS}^2 = S^2 - \langle \sigma_i^2 \rangle .
\end{equation}

If $\sigma_{\rm XS}^2 > 0$, the intrinsic variability amplitude is defined as

\begin{equation}
\mathrm{A_{\rm var}} = \sqrt{\sigma_{\rm XS}^2}.
\label{eq:Avar}
\end{equation}

Otherwise, $\mathrm{A_{\rm var}}$ is set to zero.

\subsection{Structure Function Analysis}

To characterize how variability amplitude evolves with timescale, we compute the first-order structure function (SF), a widely used diagnostic in AGN time-domain studies \citep{1985ApJ...296...46S}. The SF provides a non-parametric description of stochastic variability by measuring the typical magnitude difference between pairs of observations separated by a given time lag. It has been extensively applied to quasar variability analysis \citep[e.g.,][]{2004ApJ...601..692V,2016ApJ...817..119K,2017ApJ...849..110S,2023ApJ...958..135S}.

The noise-corrected structure function is computed as

\begin{equation}
\begin{aligned}
{\rm SF}^2(\Delta t) = \frac{1}{N_{\Delta t}}
\sum_{k=1}^{N_{\Delta t}}
\Big[
&\left(m(t_k)-m(t_k+\Delta t)\right)^2 \\
&-\left(\sigma_k^2+\sigma_{k+\Delta t}^2\right)
\Big].
\end{aligned}
\end{equation}

where $N_{\Delta t}$ is the number of magnitude pairs within a given time-lag bin $\Delta t$, and the second term removes the contribution of photometric uncertainties. Since redshift information is unavailable for a significant fraction of our sample, all time lags are computed in the observed frame, except for the SF analysis presented in section \ref{sec:var-mbh} for SDSS AGNs with redshift measurements. While rest-frame corrections are commonly applied in optical quasar variability studies, the observed-frame analysis remains appropriate for characterizing ensemble variability trends, particularly given the broad redshift range (as large as $z = 6.51$; \citealt{leung2026discoveryquasarvariabilityearly}) expected for WISE-selected AGNs. We therefore interpret the derived timescales as observed-frame quantities. Alternative robust estimators based on the interquartile range have been proposed for SF computation to reduce sensitivity to outliers \citep{2016ApJ...817..119K}. However, the variance-based definition adopted here enables direct comparison with stochastic variability models and previous MIR variability studies \citep[e.g.,][]{2017ApJ...849..110S,2023ApJ...958..135S}. 

To investigate population-level variability trends, we construct ensemble structure functions for different AGN subsamples based on source classification and physical properties. Ensemble SFs are obtained by combining all valid magnitude pairs within each time-lag bin across the subsample. Although ensemble structure functions are often approximated by a single power-law relation, particularly for sparsely sampled light curves \citep[e.g.,][]{2016ApJ...817..119K}, recent studies have demonstrated that a single slope does not always provide an adequate description over the full range of timescales \citep[e.g.,][]{2023ApJ...958..135S}.  We therefore fit the ensemble SFs using both a single power-law and a broken power-law (BPL) model. The single power-law model is expressed as

\begin{equation}
{\rm SF}(\Delta t) = A_{\rm PL} \left(\frac{\Delta t}{365}\right)^{\gamma_{\rm PL}},
\end{equation}

whereas the BPL model is given by

\begin{equation}
{\rm SF}(\Delta t) =
\begin{cases}
A \left( \dfrac{\Delta t}{365} \right)^{\gamma_1}, 
& \Delta t \le t_{\rm break} \\
\\
A \left( \dfrac{t_{\rm break}}{365} \right)^{\gamma_1}
\left( \dfrac{\Delta t}{t_{\rm break}} \right)^{\gamma_2},
& \Delta t > t_{\rm break},
\end{cases}
\end{equation}

where $\Delta t$ is the observed-frame time lag in days, $A$ represents the variability amplitude at a timescale of one year, $t_{\rm break}$ denotes the characteristic break timescale, $\gamma_1$ describes the slope at short time lags, and $\gamma_2$ characterizes the slope beyond the break. We calculate the ensemble structure function using time-lag bins of 180 days, with bin edges spanning 90--3700 days, requiring a minimum of 10 pairs in each time-lag bin to ensure a reliable estimate of the SF. No clipping or additional outlier rejection is applied to the pairwise measurements. The two SF models are compared using the Bayesian Information Criterion (BIC), defined as:

\begin{equation}
{\rm BIC} = k\ln(n) - 2\ln(\hat{\mathcal{L}}),
\end{equation}

where (k) is the number of free parameters, (n) is the number of SF data points used in the fit, and ($\hat{\mathcal{L}}$) is the maximum likelihood. A lower BIC indicates a preferred model after accounting for model complexity. Our comparison shows that the broken power-law model is preferred over the single power-law model in most of the cases considered. We therefore use the broken power-law fits to characterize the ensemble SFs and investigate the dependence of their amplitudes and slopes on AGN classification and physical properties.

For comparative analysis between different subsamples, we focus primarily on the variability amplitude at one year ($A$) and the short-timescale slope ($\gamma_1$), as these parameters are less sensitive to long-timescale sampling limitations and provide a robust characterization of the stochastic variability growth. The uncertainties on the fitted parameters are estimated using a bootstrap resampling procedure. For each subsample, we generate multiple realizations of the ensemble SF within its 1$\sigma$ uncertainties and refit the model. The dispersion of the resulting parameter distributions is adopted as the 1$\sigma$ confidence interval.

\begin{figure}
    \centering
    \includegraphics[width=9cm, height=9cm]{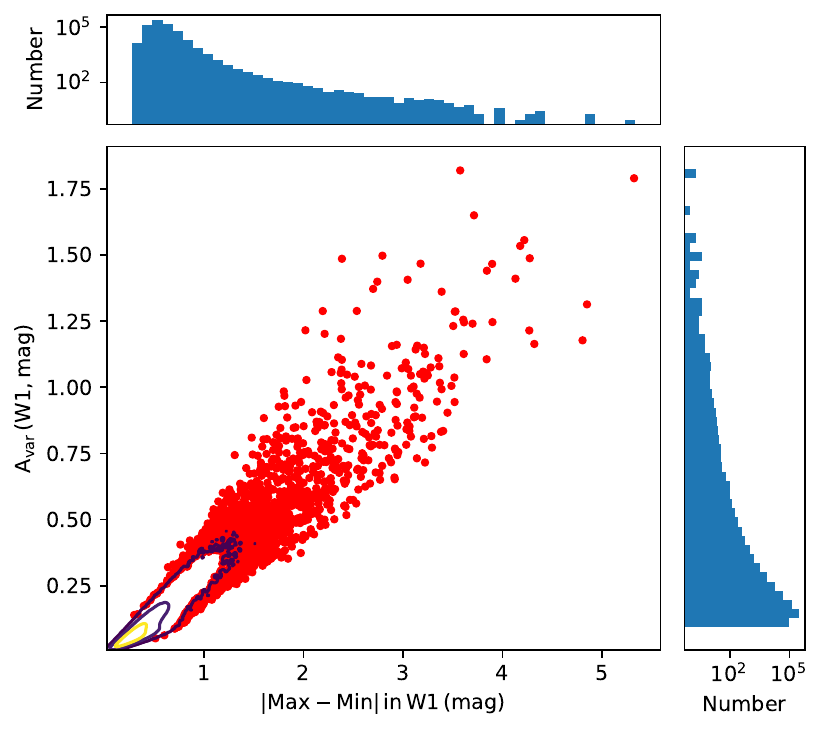}
    \caption{Variability amplitude ($\mathrm{A_{var}}$) in W1 is plotted against the maximum to minimum magnitude variation. The 1$\sigma$, 2$\sigma$, and 3$\sigma$ density contours are shown along with the outliers, which represent the objects with high variability. The corresponding histograms of the variability amplitudes are shown in the upper-left and lower-right in log-scale.}
    \label{fig:var_delmag}
\end{figure} 

\begin{figure*}
    \centering
    \includegraphics[width=15cm, height=9cm]{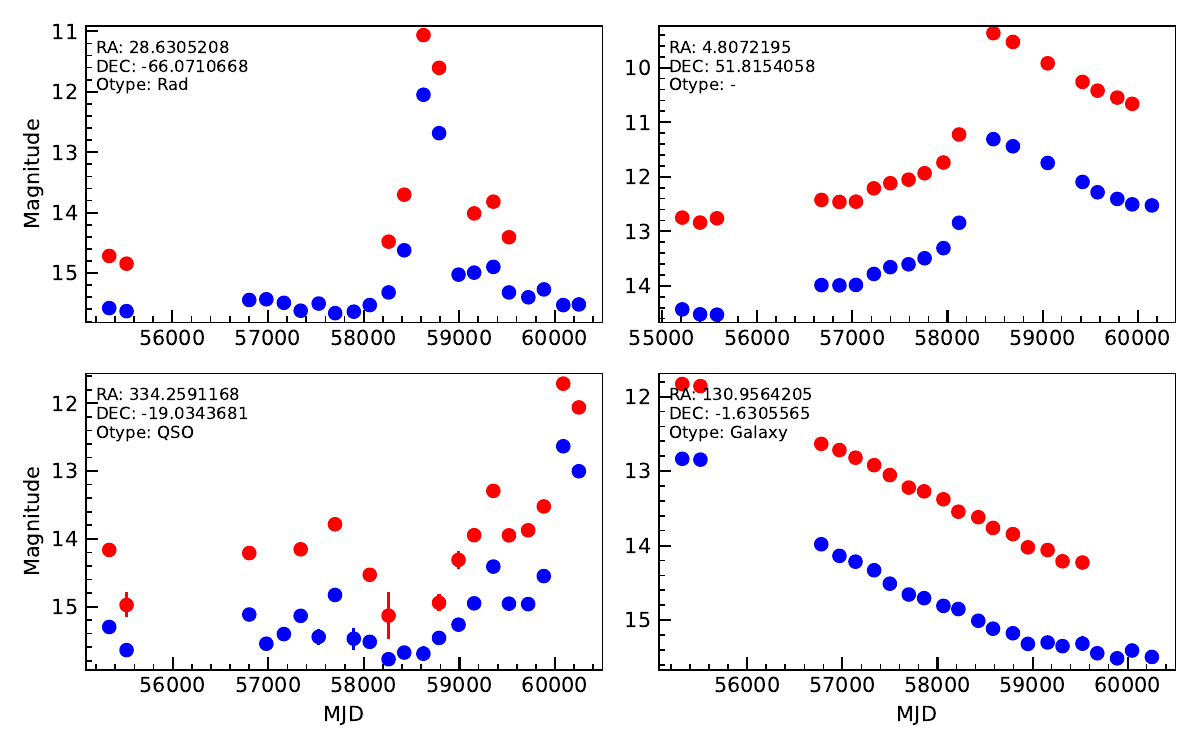}
    \caption{Example lightcurves of some variable objects in our sample in W1 (blue) and W2 (red) bands. The Object type from SIMBAD ('otype') is mentioned for each object.}
    \label{fig:lc_paper}
\end{figure*}

\section{Results}\label{sec:result}

\subsection{Variable AGNs selection}
We considered a source to be variable if $P_{\rm var}>0.95$ and $A_{\rm var}>0$, yielding a sample of 632,939 AGNs. In Figure \ref{fig:var_delmag}, we plotted the $A_{\rm var}$ against the maximum to minimum magnitude ($\Delta W1$) in the W1 band, along with density contours and the 3$\sigma$ outliers representing highly variable sources. We identify 3008 sources with $A_{\rm var} > 0.3$ mag and 624 sources with $A_{\rm var} > 0.5$ mag variation, respectively. There are 54,897 sources showing $\Delta W1>0.5$ mag in their W1 band lightcurve. The highly variable ($A_{\mathrm{var}}>0.5$ mag) sources show $\Delta W1$ variation of 1 to 5 mag within a time scale of a few years. A few examples of such highly variable sources are shown in Figure \ref{fig:lc_paper}. We note that many of our sources have no spectroscopic information, and some have very little information in different databases. The object type as found in SIMBAD is also mentioned. The lightcurves show strong variability in W1 and W2 bands, including a source showing a huge flare and classified as a radio galaxy (top-left panel), a {flare-like lightcurve with a fast rise and slow decay} (top-right), a QSO-type light curve which is in a bright phase (bottom-left), and a galaxy showing gradual dimming since the start of WISE observations (bottom-right). A detailed investigation of extreme MIR variables will be presented in a forthcoming paper. In this work, we focus on the variability analysis of the full sample and compile the derived variability properties into Table~\ref{tab:catalog}, which is made publicly available to facilitate a wide range of studies. The catalog includes a notable source, J043947.08+163415.7, which has recently been reported as a highly variable quasar at cosmic dawn ($z = 6.51$; \citealt{leung2026discoveryquasarvariabilityearly}), consistent with our findings.

\begin{deluxetable}{ll}
\tablecaption{Description of the columns in the WISE-VAULT catalog. The full catalog is available in FITS format on Zenodo under an open-source 
Creative Commons Attribution license: \dataset[doi: 10.5281/zenodo.23682572]{https://doi.org/10.5281/zenodo.23682572}.
\label{tab:catalog}}
\tablehead{
\colhead{Column Name} & \colhead{Description}
}
\startdata
Object Name & Unique source identifier \\
RA & Right ascension (J2000) \\
DEC & Declination (J2000) \\
np\_W1 & Number of data points in the light curve (W1 band) \\
np\_W2 & Number of data points in the light curve (W2 band) \\
mean\_mag\_W1 & Mean magnitude in the W1 band \\
err\_mag\_W1 & mean square uncertainty in W1 band lightcurve (eq. \ref{eq:error}) \\
del\_mag\_W1 & maximum--minimum magnitude variation in W1 \\
var\_amp\_W1 & Intrinsic variability amplitude ($A_{\rm var, W1}$) in W1 (eq. \ref{eq:Avar})  \\
chi2\_W1  &  $\chi^2$ in W1 band lightcurve (eq. \ref{eq:chi2})  \\
Pvar\_W1    &  cumulative probability of variation in W1 (eq. \ref{eq:Pvar}) \\
Vpar\_W1    &  Variability parameter in W1 (eq. \ref{eq:Vpar})  \\
mean\_mag\_W2 & Mean magnitude in the W2 band \\
err\_mag\_W2 & mean square uncertainty in W2 band lightcurve (eq. \ref{eq:error})  \\
del\_mag\_W2 & maximum--minimum magnitude variation in W2 \\
var\_amp\_W2 & Intrinsic variability amplitude ($A_{\rm var, W2}$) in W2 (eq. \ref{eq:Avar}) \\
chi2\_W2  &  $\chi^2$ in W2 band lightcurve (eq. \ref{eq:chi2})  \\
Pvar\_W2    &  cumulative probability of variation in W2 (eq. \ref{eq:Pvar}) \\
Vpar\_W2    &  Variability parameter in W2 (eq. \ref{eq:Vpar}) \\
\enddata
\end{deluxetable}

\begin{figure}
    \centering
    \includegraphics[width=9cm, height=8cm]{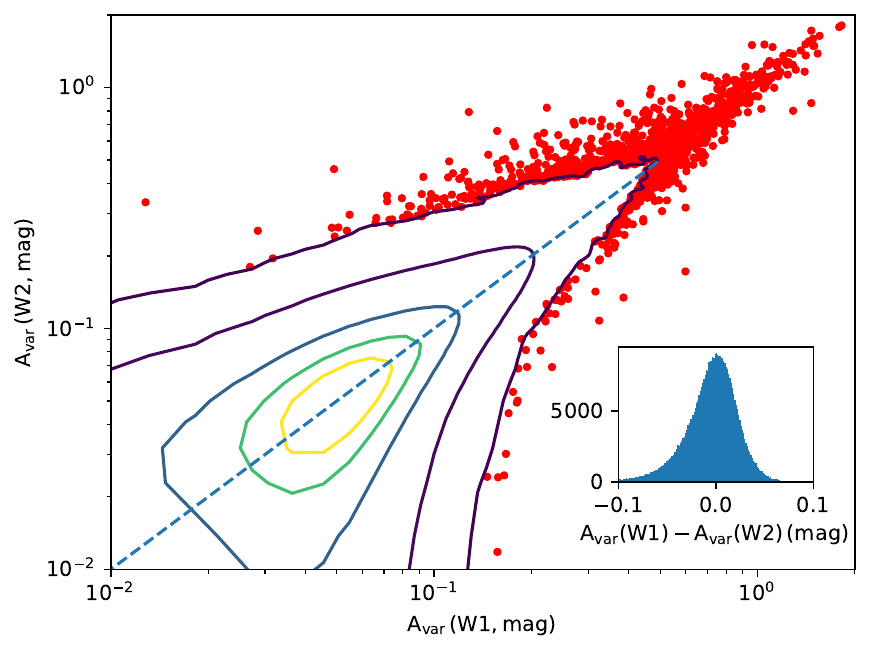}
    \caption{Comparison of variability amplitude in W1 and W2 bands showing similar correlated variation in both bands. The inset plot shows the histogram of the difference in variability amplitude in the W1 and W2 bands.}
    \label{fig:var_w1w2}
\end{figure}

We observe a strong correlation in variability between the W1 and W2 bands, as shown in Figure \ref{fig:var_w1w2} with a mean difference of 0.005 mag (a dispersion of 0.03 mag) in $A_{\mathrm{var}}$. The scatter in $A_{\mathrm{var}}$ in two bands is larger for low variable sources and reduces significantly as $A_{\mathrm{var}}$ increases, as expected. The mean W1-W2 color of the variable AGNs is 1 mag, and the majority of the sources have W1 $\sim$15 mag

It is important to note that the majority of the variable AGNs in our sample lack spectroscopic measurements. For instance, cross-matching with the SDSS DR14 quasar spectral property catalog \citep{2020ApJS..249...17R} within 3\arcsec\ search radius yields matches for only $\sim$10\% of the sources. Expanding this comparison, we cross-match our sample with the DESI quasar catalog\footnote{\url{https://data.desi.lbl.gov/public/dr1/vac/dr1/agnqso/v1.0/}} using a 3\arcsec\ search radius and identify 134,956 counterparts, corresponding to $\sim$20\% of the sample. However, the DESI catalog does not provide physical parameters such as bolometric luminosity or black hole mass. Consequently, DESI-based measurements are not included in our analysis to study the dependency of variability amplitude on AGN physical properties in Section \ref{sec:var-mbh}. 

To estimate the false-positive rate associated with our variability-selection criteria, we performed Monte Carlo simulations using a representative sample of 10000 objects. For each source, we retained the actual WISE sampling and the adopted epoch uncertainties, but removed any intrinsic variability by assigning a constant magnitude to the source. We then generated 500 mock light curves by adding Gaussian noise according to the corresponding epoch uncertainties. Each simulated light curve was analyzed using the same procedure as the real data, and sources satisfying \(V_{\rm par}>1.3\) and \(A_{\rm var}>0\) were classified as variable. We find that 7.5\% of the simulated light curves satisfy these criteria, increasing from $\sim$5\% for brighter sources to $\sim$9\% for fainter sources, providing an empirical estimate of the false-positive rate expected from statistical fluctuations and the adopted uncertainty model.

\begin{figure}
    \centering
    \includegraphics[width=9cm, height=6cm]{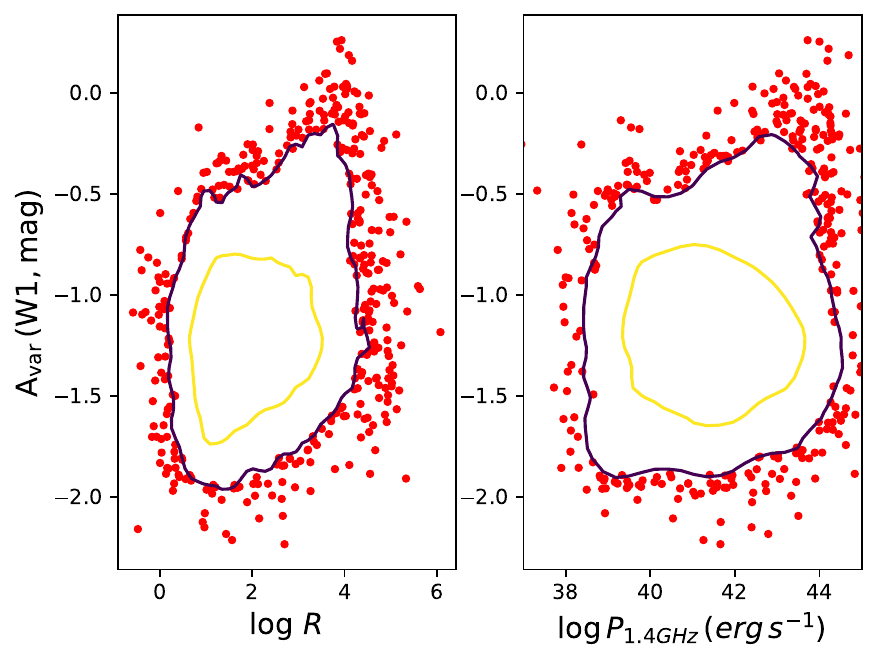}
    \caption{The correlation between variability amplitude (W1) against radio loudness (left-panel) and radio-power at 1.4GHz (right-panel) are shown. The 1$\sigma$ and 2$\sigma$ density contours are shown in thick lines along with 2$\sigma$ outliers in dots.}
    \label{fig:var_radio_corr}
\end{figure}

\begin{figure*}
    \centering
    \includegraphics[width=15cm, height=15cm]{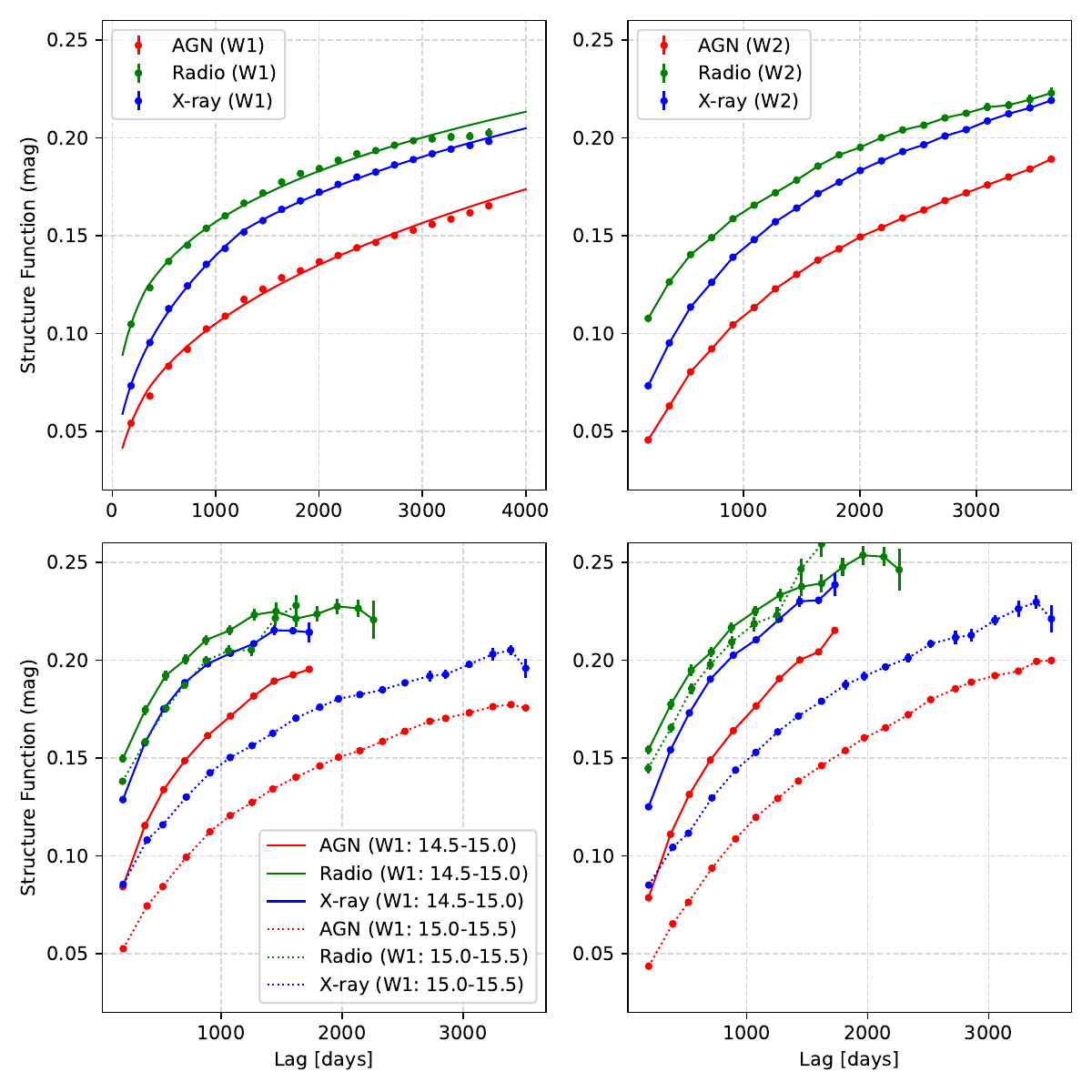}
    \caption{The ensemble structure function in the W1 band of different samples of AGN is shown. The best-fit broken power-law model is shown by solid lines. The same for W2 is shown on the right. The lower panels show the SF when samples are subdivided into smaller bins of W1 band magnitude (14.5-15.0 \& 15.0-15.5 mag) for comparison.}
    \label{fig:SF_AGN_Type}
\end{figure*}

\begin{comment}
\begin{table*}
\centering
\caption{Structure Function (SF) fitting parameters in W1 and W2 bands for different samples.}\label{table:sf_agn}
\begin{tabular}{lccccc}
\hline
Sample & Band & $A$ & $t_{\rm break}$ (days) & $\gamma_1$ & $\gamma_2$ \\
\hline
Radio & W1 & $0.127 \pm 0.002$ & $312.3 \pm 19.5$ & $0.284 \pm 0.054$ & $0.220 \pm 0.008$ \\
Radio & W2 & $0.125 \pm 0.002$ & $312.8 \pm 137.1$ & $0.231 \pm 0.063$ & $0.255 \pm 0.009$ \\
X-ray & W1 & $0.105 \pm 0.001$ & $326.8 \pm 13.8$ & $0.534 \pm 0.026$ & $0.297 \pm 0.003$ \\
X-ray & W2 & $0.104 \pm 0.001$ & $303.6 \pm 11.4$ & $0.527 \pm 0.037$ & $0.344 \pm 0.004$ \\
AGN & W1 & $0.073 \pm 0.001$ & $343.0 \pm 388.0$ & $0.445 \pm 0.016$ & $0.354 \pm 0.015$ \\
AGN & W2 & $0.075 \pm 0.004$ & $295.8 \pm 483.0$ & $0.670 \pm 0.073$ & $0.433 \pm 0.017$ \\
\hline
\end{tabular}
\end{table*}
\end{comment}

\begin{table*}
\centering
\caption{Structure Function (SF) fitting parameters in W1 and W2 bands for the different AGN samples. The broken-power-law (BPL) parameters are given by $A$, $t_{\rm break}$, $\gamma_1$, and $\gamma_2$, while $A_{\rm PL}$ and $\gamma_{\rm PL}$ denote the parameters of the single-power-law (PL) model. The final two columns give the Bayesian Information Criterion (BIC) values for the BPL and PL models, respectively.}
\label{table:sf_agn}
\begin{tabular}{lccccccccc}
\hline
Sample & Band & $A$ & $t_{\rm break}$ (days) & $\gamma_1$ & $\gamma_2$ &
$A_{\rm PL}$ & $\gamma_{\rm PL}$ & BIC$_{\rm BPL}$ & BIC$_{\rm PL}$ \\
\hline
Radio & W1 &
$0.126 \pm 0.001$ &
$319.4 \pm 5.4$ &
$0.271 \pm 0.017$ &
$0.221 \pm 0.003$ &
$0.124 \pm 0.001$ &
$0.225 \pm 0.003$ &
$49$ &
$49$ \\

Radio & W2 &
$0.126 \pm 0.001$ &
$358.9 \pm 94.4$ &
$0.229 \pm 0.019$ &
$0.251 \pm 0.004$ &
$0.126 \pm 0.001$ &
$0.249 \pm 0.003$ &
$24$ &
$19$ \\

X-ray & W1 &
$0.096 \pm 0.001$ &
$1258.4 \pm 78.3$ &
$0.375 \pm 0.005$ &
$0.256 \pm 0.005$ &
$0.099 \pm 0.001$ &
$0.315 \pm 0.001$ &
$65$ &
$1304$ \\

X-ray & W2 &
$0.101 \pm 0.002$ &
$349.2 \pm 490.4$ &
$0.462 \pm 0.033$ &
$0.346 \pm 0.018$ &
$0.099 \pm 0.001$ &
$0.353 \pm 0.001$ &
$357$ &
$465$ \\

AGN & W1 &
$0.073 \pm 0.001$ &
$343.0 \pm 118.5$ &
$0.434 \pm 0.004$ &
$0.363 \pm 0.005$ &
$0.072 \pm 0.001$ &
$0.367 \pm 0.001$ &
$8138$ &
$8767$ \\

AGN & W2 &
$0.064 \pm 0.001$ &
$1382.0 \pm 11.4$ &
$0.515 \pm 0.001$ &
$0.394 \pm 0.001$ &
$0.068 \pm 0.001$ &
$0.449 \pm 0.001$ &
$390$ &
$6360$ \\
\hline
\end{tabular}
\end{table*}

\subsection{Variability of different AGN samples}\label{sec:diff_agn_sample}

In this section, we investigate the variability properties of our AGN sample in the context of their multiwavelength counterparts across different surveys.

\subsubsection{Variability of Radio-detected Sample}\label{sec:radio}
We cross-matched our variable AGN sample with the Faint Images of the Radio Sky at Twenty-Centimeters survey \citep[FIRST;][]{1997ApJ...475..479W} using a matching radius of 5\arcsec, yielding 16,406 radio-detected sources. To characterize their radio properties, we computed the radio-loudness parameter, defined as the logarithmic ratio of the radio flux density at 1.4\,GHz to the optical $g$-band flux. The $g$-band magnitudes were obtained from the SIMBAD database using a matching radius of 3\arcsec\ from the WISE positions. The resulting radio-loudness parameter spans the range $\log R = -0.6$ to $6$. Note that since the optical \(g\)-band fluxes are taken from SIMBAD, they may originate from heterogeneous surveys and epochs. Therefore, the resulting radio-loudness estimates should be regarded as approximate, and differences in optical coverage and photometric measurements may introduce additional scatter.

Following standard conventions, we classify sources with $\log R < 1$ as radio-quiet (RQ) and those with $\log R > 1$ as radio-loud (RL), yielding 7528 RL AGNs in our sample. We further estimate the radio power at 1.4\,GHz assuming a power-law spectrum of the form $f_\nu \propto \nu^{-\alpha}$ with a fixed spectral index $\alpha = 0.8$. The radio luminosity is calculated as \citep[e.g.,][]{1992ARA&A..30..575C}

\begin{equation}
P_{\rm 1.4\,GHz} = 4\pi D_L^2 \, (1+z)^{\alpha - 1} \, f_{\nu,\,1.4\, \rm GHz},
\end{equation}

where $D_L$ is the luminosity distance and $f_{\nu,\,1.4\,GHz}$ is the observed flux density at 1.4\,GHz. Note that $P_{\rm 1.4\,GHz}$ was calculated for sources having redshift.
 
In Figure \ref{fig:var_radio_corr}, we plotted the correlation of $A_{\rm{var}}$ with radio-loudness and radio-power and found a weak but statistically significant positive correlation between the variability amplitude and radio-loudness ($R$), with a Spearman rank coefficient of $\rho \approx 0.20$ and a very small p-value ($p \sim 4\times10^{-39}$). In contrast, the correlation between variability amplitude and radio power is negligible ($\rho \approx 0.04$). This indicates a modest contribution from jet-related non-thermal emission, though not the dominant driver of MIR variability \citep[see also][]{2019ApJ...886...33L,2019MNRAS.483.2362R}. This is consistent with the expectation that, for predominantly non-blazar AGNs, MIR variability arises from thermal dust reprocessing of accretion disk fluctuations rather than non-thermal jet emission. We note that the radio-loud and radio-quiet samples may differ in redshift and luminosity, and the observed variability differences should therefore not be attributed solely to jet or non-thermal emission.

\begin{figure*}
    \centering
    \includegraphics[width=15cm, height=6cm]{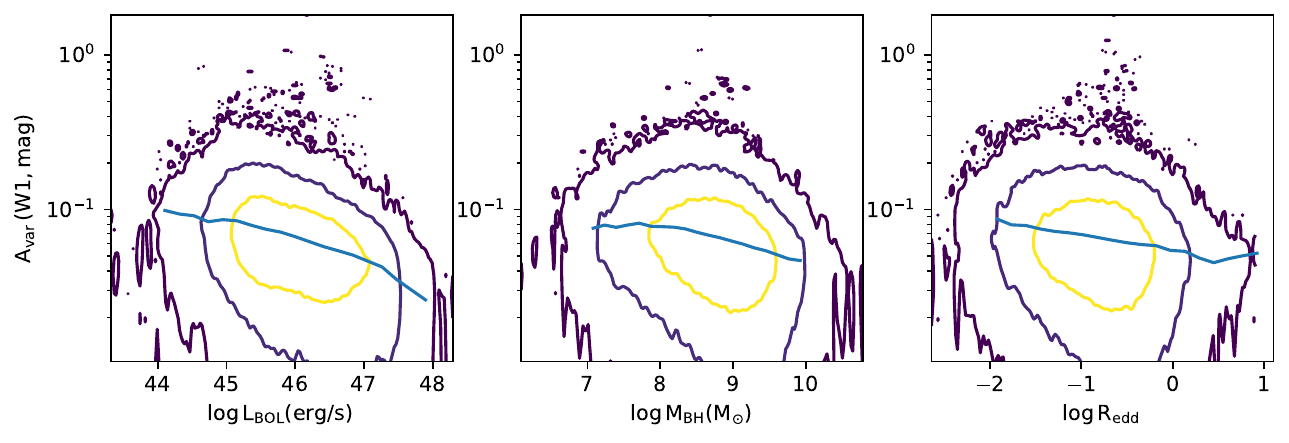}
    \caption{Correlation of variability amplitude (W1) with AGN physical parameters such as bolometric luminosity (left), black hole mass (middle), and Eddington ratio (right). The 1$\sigma$, 2$\sigma$, and 3$\sigma$ density contours are shown along with the median variability at different bins of these physical parameters.}
    \label{fig:var_corr}
\end{figure*}

\subsubsection{Variability of X-ray-detected Sample}\label{sec:x-ray}
To identify X-ray counterparts of our variable AGN sample, we cross-matched our sources with several major X-ray catalogs using a search radius of 5\arcsec, following the convention commonly used in previous studies \citep[e.g.,][]{2026ApJS..285...20P}. Specifically, we identified 5641 matches in the \textit{Chandra} Source Catalog v2.0 \citep{2010ApJS..189...37E}, 59,135 matches in the eROSITA Final Equatorial Depth Survey AGN catalog \citep[eFEDs;][]{2022A&A...661A...5L}, 13,224 matches in the \textit{XMM-Newton} serendipitous source catalog \citep[4XMM-DR12;][]{2020A&A...641A.136W}, and 18,837 matches in the \textit{Swift} X-ray Telescope Point Source Catalog \citep{2023MNRAS.518..174E}. In cases where a source is detected in multiple X-ray surveys, we retain only a single counterpart to avoid duplication. After accounting for overlapping detections, we obtain a total of 86,253 unique AGNs with X-ray counterparts. The X-ray-detected subsample is drawn from surveys with different depths and sky coverage, and therefore does not represent a uniform X-ray selection of the full AGN sample.

In Figure~\ref{fig:SF_AGN_Type}, we plotted the ensemble structure function in the W1 band of AGN samples, namely, radio-detected, X-ray-detected, and variable-AGN samples (which include all variable sources found in this work). The best-fit parameters for the broken power-law fit of the SF are given in Table~\ref{table:sf_agn}. The structure function (SF) analysis reveals clear and systematic differences in variability properties among the different AGN populations. The variability amplitude at 1 yr ($A$) shows a strong dependence on AGN type, with radio-detected sources exhibiting the highest amplitudes ($A=0.126\pm0.001$), followed by X-ray-detected sources ($A= 0.096\pm 0.001$), and the variable-AGN sample ($A=0.073\pm 0.001$). This trend indicates that radio-detected AGNs are the most variable population in the MIR, consistent with the presence of additional non-thermal components, such as jet emission, contributing to the variability. 

To investigate whether the differences in MIR variability among the radio-detected, X-ray-detected, and general AGN samples are driven by differences in brightness, we further constructed ensemble structure functions in two W1 magnitude intervals, $14.5\leq W1<15.0$ and $15.0\leq W1<15.5$, as shown in the lower panels of Figure~\ref{fig:SF_AGN_Type}. In both magnitude intervals, we recover the same relative trend, with the radio-detected AGNs exhibiting the highest variability amplitude, followed by the X-ray-detected AGNs and the general AGN sample. This persistence of the relative ordering within narrow W1 magnitude intervals indicates that the observed differences in variability are not solely driven by differences in the W1 magnitude distributions of the samples. We also find that the SF amplitude is systematically lower in the fainter W1 magnitude interval, indicating a dependence of the measured MIR variability amplitude on source brightness.

The short-timescale slopes ($\gamma_1$) also differ among the samples. X-ray detected AGNs show the steepest slopes ($\gamma_1 \sim 0.38-0.46$), implying rapid growth of variability at short timescales, consistent with strong coupling to the inner accretion disk. Radio-detected sources exhibit significantly flatter slopes ($\gamma_1 \sim 0.23$--0.27), indicating slower variability growth at short timescales. This suggests that radio AGNs have more gradual and smoother variability evolution, possibly due to the contribution of extended or jet-related emission. At longer timescales, the SF slopes ($\gamma_2$) flatten for all samples, indicating saturation of variability. However, X-ray and variable-AGN samples maintain relatively higher $\gamma_2$ values compared to radio sources, suggesting that their variability continues to evolve over longer timescales, whereas radio sources reach saturation earlier.  The value of $\gamma_1$ found in our variable-AGN sample is consistent with the finding of $\gamma_1=0.47-0.51$ obtained in local Type 1 and Type II AGNs by \citet{2023ApJ...958..135S} and the $\gamma \sim 0.45$ derived from the MIR lightcurves of quasars by \citet{2016ApJ...817..119K}.

The fitted break timescales vary across the AGN subsamples and between the W1 and W2 bands. These values should be interpreted cautiously because they may be affected by the approximately six-month cadence of WISE observations, source brightness, and photometric uncertainties. Furthermore, reliable redshifts are unavailable for a substantial fraction of the sample, so the SFs in this section are measured in the observed frame. We therefore treat ($t_{\rm break}$) as a phenomenological parameter describing the shape of the ensemble SFs rather than as a direct measurement of a physical dust-reprocessing or accretion timescale. Previous studies have shown that MIR variability and SF shapes can contain information on dust reverberation and torus structure \citep[e.g.,][]{2019ApJ...886...33L,2023ApJ...950..122L}; however, extracting such physical scales requires well-characterized redshifts and complementary well-sampled optical/MIR data.

%This indicates that the dominant MIR variability mechanism operates on similar physical scales across these populations, likely associated with dust reprocessing in the circumnuclear torus or characteristic accretion-driven fluctuations. 

\begin{figure}
    \centering
    \includegraphics[width=9cm, height=15cm]{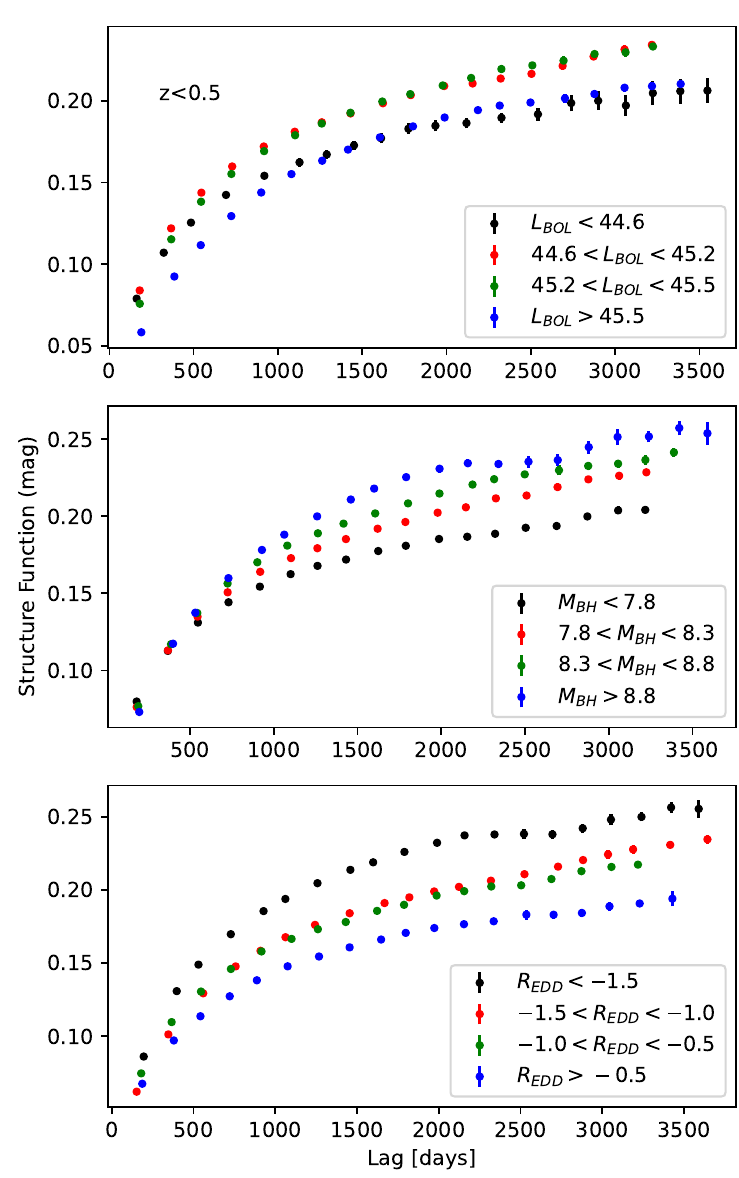}
    \caption{The ensemble structure function in W1 band for the SDSS-DR14 quasars for different bins of $L_{\rm BOL}$, $M_{\rm BH}$, and $R_{\rm EDD}$ at $z<0.5$.}
    \label{fig:SF_AGN_par}
\end{figure}

\subsection{Correlation of variability amplitude with AGN parameters}\label{sec:var-mbh}

To investigate the dependence of infrared variability on fundamental AGN physical parameters, we examine correlations between the variability amplitude ($A_{\rm var}$) and the bolometric luminosity ($L_{\rm bol}$), black hole mass ($M_{\rm BH}$), and Eddington ratio ($R_{\rm Edd}$) adopted from the catalog of \citet{2020ApJS..249...17R}. In this catalog, bolometric luminosities are estimated from monochromatic continuum luminosities at 5100\,\AA, 3000\,\AA, and 1350\,\AA\, depending on the redshift, using appropriate bolometric corrections. Black hole masses are derived using single-epoch virial estimators based on broad emission lines, specifically H$\beta$, Mg\,II, and C\,IV, depending on the redshift of the source. The Eddington ratio is defined as $R_{\rm Edd} = L_{\rm bol}/L_{\rm Edd}$, where the Eddington luminosity $L_{\rm Edd}$ is computed from the black hole mass. Further details of these measurements are provided in \citet{2020ApJS..249...17R}.

In Figure~\ref{fig:var_corr}, we present the correlations between $A_{\rm var}$ and $L_{\rm bol}$, $M_{\rm BH}$, and $R_{\rm Edd}$. We find that the variability amplitude exhibits a systematic anti-correlation with all three parameters, with Spearman rank coefficients of $\rho \approx -0.38$, $-0.27$, and $-0.20$ for $L_{\rm bol}$, $M_{\rm BH}$, and $R_{\rm Edd}$, respectively, all with highly significant (very small) $p$-values. As a robustness check, we examined the correlations between the variability amplitude and the AGN physical parameters within bins of the other relevant parameters (see Table \ref{tab:sf_z1} for bin size). The anti-correlation between $A_{\rm var}$ and $L_{\rm bol}$ persists when the sample is divided into four $M_{\rm BH}$ bins, with $\rho$ ranging from $\rho=-0.15$ to $-0.39$. Similarly, the correlation between $A_{\rm var}$ and $M_{\rm BH}$ was examined within four $L_{\rm bol}$ bins, yielding $\rho=0.03$--$0.18$ in three bins and $\rho=-0.21$ in the highest-$L_{\rm bol}$ bin, suggesting the dependence on $M_{\rm BH}$ is comparatively weak and changes across the $L_{\rm bol}$ bins. The anti-correlation between $A_{\rm var}$ and $R_{\rm Edd}$ also persists within the four $L_{\rm bol}$ bins, with $\rho=-0.04$ to $-0.17$. Such trends are broadly consistent with previous optical variability studies, which have reported that AGNs with higher luminosities and accretion rates tend to exhibit lower variability amplitudes \citep[e.g.,][]{2004ApJ...601..692V,2010ApJ...721.1014M,2017ApJ...842...96R}. 

In order to further investigate the dependence of variability on AGN physical parameters, we construct ensemble structure functions in bins of $L_{\rm bol}$, $M_{\rm BH}$, and $R_{\rm Edd}$ for $z<0.5$. Here, the structure functions are calculated by converting the time lags to the rest frame using $\tau_{\rm rest}=\tau_{\rm obs}/(1+z)$. The resulting rest-frame ensemble SFs are shown in Figure \ref{fig:SF_AGN_par}, and the corresponding best-fit parameters are summarized in Table~\ref{tab:sf_z1}. We find clear systematic trends of variability amplitude with AGN properties. The SF amplitude ($A$) decreases with increasing bolometric luminosity, indicating that more luminous AGNs exhibit reduced variability across all timescales. A weaker but noticeable dependence is also observed with black hole mass, where higher-mass systems tend to show slightly lower variability amplitudes.

The strongest dependence is seen with the Eddington ratio. AGNs with lower $R_{\rm Edd}$ display significantly higher variability amplitudes compared to high accretion rate systems, indicating that variability is closely linked to the accretion state. This trend is consistently reflected in the fitted SF amplitudes listed in Table~\ref{tab:sf_z1}, where $A$ decreases systematically toward higher $R_{\rm Edd}$. The structure function slopes further support this interpretation. The short-timescale slope ($\gamma_1$) typically ranges between $\sim0.4$--$0.6$, while the long-timescale slope ($\gamma_2$) is flatter ($\sim0.2$--$0.4$), indicating a gradual saturation of variability at longer timescales. The break timescale ($t_{\rm break}$) is generally of order $\sim500$--$600$ days, with a tendency to increase for higher luminosity sources.

%suggesting longer characteristic variability timescales in more stable accretion systems.

Despite these differences in amplitude, the overall shape of the structure function remains broadly similar across different parameter bins, implying that the underlying stochastic variability process is largely universal. Instead, the normalization of the variability appears to be primarily regulated by the physical properties of the accretion flow. These results demonstrate that MIR variability is most strongly governed by the Eddington ratio, with secondary dependence on luminosity and black hole mass.

\begin{table*}
\centering
\caption{Broken-power-law structure-function fitting parameters for AGN subsamples at $z<0.5$, divided according to bolometric luminosity ($L_{\rm bol}$), black-hole mass ($M_{\rm BH}$), and Eddington ratio ($R_{\rm Edd}$). The broken power-law model is preferred over the single power-law model for all subsamples considered here.}
\label{tab:sf_z1}
\begin{tabular}{lccccc}
\hline
Parameter bin & Band & $A$ & $t_{\rm break}$ (days) & $\gamma_1$ & $\gamma_2$ \\
\hline

\multicolumn{6}{c}{$L_{\rm bol}$} \\
\hline
$L_{\rm bol}<44.6$ & W1 &
$0.112\pm0.002$ & $640.8\pm139.6$ &
$0.423\pm0.037$ & $0.229\pm0.016$ \\

 & W2 &
$0.115\pm0.003$ & $479.4\pm146.3$ &
$0.554\pm0.052$ & $0.335\pm0.016$ \\

$44.6<L_{\rm bol}<45.2$ & W1 &
$0.122\pm0.001$ & $501.3\pm16.9$ &
$0.525\pm0.009$ & $0.271\pm0.003$ \\

 & W2 &
$0.114\pm0.001$ & $498.6\pm30.1$ &
$0.590\pm0.013$ & $0.348\pm0.005$ \\

$45.2<L_{\rm bol}<45.5$ & W1 &
$0.115\pm0.001$ & $508.8\pm34.3$ &
$0.589\pm0.015$ & $0.300\pm0.006$ \\

 & W2 &
$0.100\pm0.001$ & $527.2\pm84.0$ &
$0.621\pm0.025$ & $0.378\pm0.011$ \\

$L_{\rm bol}>45.5$ & W1 &
$0.088\pm0.001$ & $872.2\pm108.9$ &
$0.587\pm0.027$ & $0.297\pm0.021$ \\

 & W2 &
$0.073\pm0.001$ & $1024.3\pm102.5$ &
$0.587\pm0.018$ & $0.367\pm0.014$ \\

\hline
\multicolumn{6}{c}{$M_{\rm BH}$} \\
\hline
$M_{\rm BH}<7.8$ & W1 &
$0.112\pm0.001$ & $510.2\pm62.5$ &
$0.485\pm0.020$ & $0.246\pm0.008$ \\

 & W2 &
$0.101\pm0.001$ & $499.5\pm83.2$ &
$0.525\pm0.026$ & $0.326\pm0.007$ \\

$7.8<M_{\rm BH}<8.3$ & W1 &
$0.113\pm0.001$ & $507.7\pm27.1$ &
$0.555\pm0.011$ & $0.298\pm0.005$ \\

 & W2 &
$0.099\pm0.001$ & $503.6\pm46.8$ &
$0.607\pm0.017$ & $0.379\pm0.006$ \\

$8.3<M_{\rm BH}<8.8$ & W1 &
$0.113\pm0.001$ & $531.3\pm69.2$ &
$0.594\pm0.021$ & $0.314\pm0.013$ \\

 & W2 &
$0.100\pm0.001$ & $711.5\pm116.2$ &
$0.593\pm0.027$ & $0.360\pm0.018$ \\

$M_{\rm BH}>8.8$ & W1 &
$0.109\pm0.001$ & $665.9\pm25.5$ &
$0.641\pm0.011$ & $0.312\pm0.006$ \\

 & W2 &
$0.097\pm0.001$ & $701.7\pm94.5$ &
$0.674\pm0.019$ & $0.388\pm0.018$ \\

\hline
\multicolumn{6}{c}{$R_{\rm Edd}$} \\
\hline
$R_{\rm Edd}<-1.5$ & W1 &
$0.122\pm0.001$ & $649.9\pm18.4$ &
$0.558\pm0.009$ & $0.270\pm0.004$ \\

 & W2 &
$0.116\pm0.001$ & $655.1\pm30.1$ &
$0.614\pm0.012$ & $0.345\pm0.005$ \\

$-1.5<R_{\rm Edd}<-1.0$ & W1 &
$0.104\pm0.001$ & $547.1\pm65.9$ &
$0.597\pm0.018$ & $0.315\pm0.012$ \\

 & W2 &
$0.091\pm0.001$ & $731.1\pm71.0$ &
$0.605\pm0.018$ & $0.364\pm0.010$ \\

$-1.0<R_{\rm Edd}<-0.5$ & W1 &
$0.109\pm0.001$ & $516.0\pm52.5$ &
$0.543\pm0.017$ & $0.289\pm0.008$ \\

 & W2 &
$0.094\pm0.001$ & $526.7\pm96.0$ &
$0.575\pm0.023$ & $0.371\pm0.011$ \\

$R_{\rm Edd}>-0.5$ & W1 &
$0.094\pm0.001$ & $719.8\pm156.7$ &
$0.488\pm0.032$ & $0.271\pm0.021$ \\

 & W2 &
$0.080\pm0.001$ & $1012.1\pm224.5$ &
$0.475\pm0.035$ & $0.337\pm0.018$ \\

\hline
\end{tabular}
\end{table*}

\section{Discussions}\label{sec:discussion}
\subsection{Highly Variable AGNs}

A subset of our sample exhibits extreme MIR variability, with 54,897 AGNs showing maximum--minimum variations of $\Delta W1 > 0.5$ mag. In addition, 624 sources display large intrinsic variability amplitudes with $A_{\rm var} > 0.5$ mag. Such high-amplitude variability is uncommon for typical stochastic AGN light curves and points to substantial changes in the accretion flow or its reprocessing environment. A large fraction of these highly variable sources are radio-undetected, suggesting that their variability is unlikely to be dominated by jet-related processes. These extreme MIR-variable sources provide a promising sample for investigating rare and rapidly evolving nuclear phenomena. Some of their variability patterns may be consistent with those expected from changing-look AGNs (CL-AGNs) or tidal disruption events (TDEs), but MIR variability alone is insufficient to establish such classifications. 

In the CL-AGN scenario, large-amplitude MIR variations are interpreted as dust echoes of significant changes in the accretion rate, often accompanied by the appearance or disappearance of broad emission lines \citep{2017ApJ...846L...7S}. We cross-matched our catalog with the recent homogeneous catalog of known changing-look AGNs (CL-AGNs) compiled by \citet{camus2026compilinglargestmultiwavelengthdatabase}. Their catalog contains 1,438 unique sources, including 987 spectroscopically confirmed CL-AGNs and 451 candidate or photometrically selected systems \citep{camus2026compilinglargestmultiwavelengthdatabase}. We find 533 of these known CL-AGNs in our catalog. The MIR variability amplitudes of these CL-AGNs in the W1 band have a median of $\sim0.15$ mag, with the most variable source showing $\sim0.67$ mag.

TDEs can produce luminous ultraviolet/optical flares that heat the surrounding dust, leading to delayed and long-lived infrared emission \citep{2016ApJ...829...19V,2021ApJS..252...32J}. To place our sample in the context of previously identified infrared transients, we cross-match with catalogs of known MIR outbursts. We identify 7 matches with the ``MIRONG'' sample \citep{2021ApJS..252...32J}, which consists of nearby galaxies exhibiting large-amplitude infrared flares associated with extreme accretion events. We also find 122 matches with the ``Flaires'' catalog \citep{2025A&A...695A.228N}, which compiles a large population of infrared flares interpreted as dust echoes of transient accretion episodes. We also cross-matched our catalog with the \citet{2026ApJ...999..243F} TDE catalog from the `Open mulTiwavelength Transient Event Repository' (OTTER), which provides a multiwavelength compilation of published TDEs. The initial OTTER release contains photometric data for 232 TDEs spanning wavelengths from radio to X-rays \citep{2026ApJ...999..243F}. We identify 9 matches between the OTTER TDE catalog and ours. Their MIR variability amplitudes in the W1 band have a median of $\sim0.22$ mag, with the most variable source showing $\sim0.35$ mag. These associations demonstrate that our variability selection recovers sources previously identified as infrared transients and TDEs, highlighting the potential of long-term MIR variability searches for identifying nuclear transient phenomena.
%The presence of these matches supports the interpretation that a subset of our highly variable sources is associated with CL-AGN or TDE-like events. 
The large number of high-amplitude MIR variables identified in this work highlights the potential of WISE time-domain data for uncovering rare and rapidly evolving accretion phenomena. The sources with $\Delta W1 > 0.5$ mag and $A_{\rm var} > 0.5$ mag represent a particularly compelling set of candidates for future spectroscopic and multi-wavelength follow-up aimed at identifying changing-look behavior and transient accretion events.

In this context, upcoming large time-domain surveys such as the Vera C. Rubin Observatory Legacy Survey of Space and Time \citep[LSST;][]{2019ApJ...873..111I} will be particularly valuable, providing dense optical monitoring that complements the MIR variability and enables a more complete understanding of the underlying accretion-driven processes. In addition, space missions such as SPHEREx \citep{2014arXiv1412.4872D} and \textit{Euclid} \citep{2022A&A...662A.112E} will provide complementary near-infrared spectro-photometric coverage and high-resolution imaging, enabling detailed characterization of dust emission, robust redshift determination, and AGN--host decomposition, thereby offering critical context for understanding the physical origin and environment of the observed variability.

\subsection{Wavelength-dependent Variability}

\begin{figure}
    \centering
    \includegraphics[width=9cm, height=12cm]{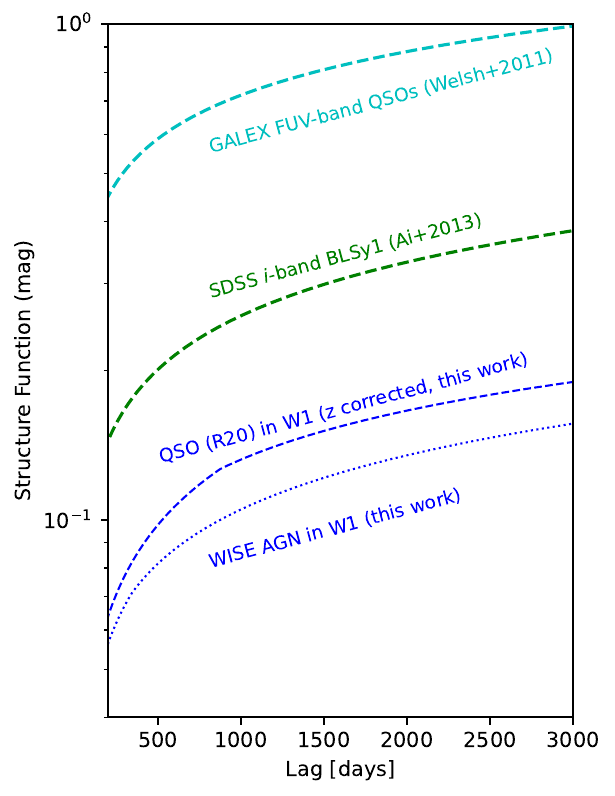}
    \caption{Comparison of the ensemble structure function of quasars at different wavelength regions, GALEX FUV-band quasars from \citet{2011A&A...527A..15W}, SDSS i-band broad-line AGN from \citet{2013AJ....145...90A}, SDSS DR14 quasars (SF corrected for redshift) and WISE all variable AGNs in W1 bands (SF uncorrected for redshift).}
    \label{fig:SF_wavelength}
\end{figure}

The dependence of AGN variability on wavelength provides important insight into the physical origin of the emission regions. In Figure~\ref{fig:SF_wavelength}, we compare the structure function derived for SDSS DR14 quasars after correcting for redshift (see section \ref{sec:var-mbh}) and all variable AGNs (redshift uncorrected) obtained in this work from WISE W1-band light curves with those obtained at shorter wavelengths, including the GALEX FUV band \citep{2011A&A...527A..15W} and the SDSS $i$-band \citep{2013AJ....145...90A}. As shown in the figure, the variability amplitude exhibits a clear wavelength dependence, with the largest amplitudes observed in the ultraviolet, intermediate amplitudes in the optical, and the lowest amplitudes in the MIR.

At a given timescale, the UV structure function lies significantly above the optical and infrared curves, while the WISE W1-band variability is systematically suppressed across all time lags. This trend is consistent with numerous previous studies, which have shown that AGN variability amplitude decreases with increasing wavelength \citep[e.g.,][]{2004ApJ...601..692V,2010ApJ...721.1014M,2016ApJ...817..119K}. The enhanced variability at shorter wavelengths is generally attributed to emission originating from the inner regions of the accretion disk, where temperature fluctuations and accretion rate variations are more pronounced.

In contrast, MIR emission arises predominantly from circumnuclear dust that reprocesses the variable ultraviolet and optical radiation from the accretion disk \citep{1987ApJ...320..537B,2019ApJ...886...33L}. This reprocessing introduces both temporal smoothing and dilution of variability amplitude, leading to a reduced structure function normalization in the infrared. The similarity in the overall shape of the structure function across wavelengths suggests that the underlying driving mechanism is common, while the amplitude modulation reflects the spatial extent and thermal response of the emitting region. This wavelength dependence is consistent with previous studies of WISE AGNs. \citet{2019ApJ...886...33L} found that the MIR variability amplitude generally decreases toward longer wavelengths, with blazars and flat-spectrum radio sources showing an exception to this trend, likely due to the contribution from non-thermal jet emission. Our finding of stronger variability in W1 than in W2 is broadly consistent with this wavelength-dependent behavior. These results support a picture in which AGN variability originates in the accretion disk and propagates outward, with the infrared variability tracing a delayed and smoothed response of the dusty torus. The wavelength dependence of the structure function is consistent with the expected connection between accretion-disk variability and its reprocessed emission from circumnuclear dust.

\section{Conclusion}\label{sec:conclusion}

We present a comprehensive study of MIR variability in a large sample of 4,543,530 WISE color-selected AGNs, aimed at characterizing their variability properties and identifying extreme MIR-variable sources that provide promising targets for follow-up studies of changing-look AGNs (CL-AGNs) and tidal disruption events (TDEs). Using multi-epoch photometry from \textit{WISE} and NEOWISE spanning over a decade, we quantify variability through maximum--minimum amplitudes, excess variance, reduced $\chi^2$, and structure function (SF) analysis over timescales of 0.5–10 years. Our main findings are summarized below:

\begin{itemize}

\item We identify 632,939 variable AGNs, including 54,897 sources exhibiting large maximum-to-minimum variability ($\Delta W1 > 0.5$ mag) and 624 sources with extreme intrinsic variability ($A_{\rm var} > 0.5$ mag). A subset of these sources shows flare-like behavior, making them promising candidates for CL-AGN or TDE events and warranting spectroscopic and multi-wavelength follow-up. The variability in the W1 and W2 bands is strongly correlated.

\item The SF analysis shows that radio-detected AGNs exhibit the highest variability amplitudes (A=$0.127\pm0.001$ mag), followed by X-ray-detected sources (A=$0.096\pm0.001$ mag) and the overall AGN sample (A=$0.073\pm0.001$ mag). Within the radio-detected subsample, the variability amplitude shows a positive correlation with radio-loudness, suggesting a contribution from non-thermal synchrotron emission in addition to thermal emission from the accretion disk.

\item The broken power-law fits to the SF further reveal that X-ray-detected AGNs exhibit steeper short-timescale slopes ($\gamma_1\approx0.38 - 0.46$), indicating rapid growth of variability, whereas radio-detected sources show flatter slopes ($\gamma_1 \sim 0.23 - 0.27$), consistent with smoother variability evolution. At longer timescales, the SF slopes flatten for all samples, indicating saturation of variability. These differences suggest that the ensemble MIR variability properties depend on AGN population and emission characteristics.

\item By examining the dependence of variability on physical parameters, we find that MIR variability is strongly anti-correlated with Eddington ratio, such that high-accretion-rate sources exhibit lower variability amplitudes compared to low-accretion-rate systems. This trend is consistent with results from optical studies and supports a scenario in which MIR variability primarily arises from dust reprocessing of the UV/optical emission from the accretion disk. The ensemble SFs generally exhibit steeper slopes at shorter time lags ($\gamma_1$) than at longer time lags ($\gamma_2$), indicating a change in the temporal scaling of MIR variability across timescales.
\end{itemize}

\begin{acknowledgments}
We thank the Reviewer for their constructive comments and suggestions, which have helped improve the quality and clarity of the manuscript. This publication makes use of data products from the Wide-field Infrared Survey Explorer, which is a joint project of the University of California, Los Angeles, and the Jet Propulsion Laboratory/California Institute of Technology, funded by the National Aeronautics and Space Administration. 
\end{acknowledgments}

\vspace{5mm}
\facilities{IRSA}
\software{TOPCAT \citep{2005ASPC..347...29T}, NumPy \citep{numpy}, SciPy \citep{2020NatMe..17..261V}, Matplotlib \citep{Matplotlib}, Astropy \citep{astropy}.}

%\bibliography{biblio}
\bibliography{cleaned_biblio}

@ARTICLE{2013AJ....145...90A,
       author = {{Ai}, Y.~L. and {Yuan}, W. and {Zhou}, H. and {Wang}, T.~G. and {Dong}, X.-B. and {Wang}, J.~G. and {Lu}, H.~L.},
        title = "{A Comparative Study of Optical/Ultraviolet Variability of Narrow-line Seyfert 1 and Broad-line Seyfert 1 Active Galactic Nuclei}",
      journal = {\aj},
         year = 2013,
        month = apr,
       volume = {145},
       number = {4},
          eid = {90},
        pages = {90},
          doi = {10.1088/0004-6256/145/4/90},
archivePrefix = {arXiv},
       eprint = {1301.4739},
 primaryClass = {astro-ph.CO},
       adsurl = {https://ui.adsabs.harvard.edu/abs/2013AJ....145...90A}
}

@ARTICLE{1993ARA&A..31..473A,
       author = {{Antonucci}, Robert},
        title = "{Unified models for active galactic nuclei and quasars.}",
      journal = {\araa},
         year = 1993,
        month = jan,
       volume = {31},
        pages = {473-521},
          doi = {10.1146/annurev.aa.31.090193.002353},
       adsurl = {https://ui.adsabs.harvard.edu/abs/1993ARA&A..31..473A}
}

@ARTICLE{2018ApJS..234...23A,
       author = {{Assef}, R.~J. and {Stern}, D. and {Noirot}, G. and {Jun}, H.~D. and {Cutri}, R.~M. and {Eisenhardt}, P.~R.~M.},
        title = "{The WISE AGN Catalog}",
      journal = {\apjs},
         year = 2018,
        month = feb,
       volume = {234},
       number = {2},
          eid = {23},
        pages = {23},
          doi = {10.3847/1538-4365/aaa00a},
archivePrefix = {arXiv},
       eprint = {1706.09901},
 primaryClass = {astro-ph.GA},
       adsurl = {https://ui.adsabs.harvard.edu/abs/2018ApJS..234...23A}
}

@ARTICLE{1987ApJ...320..537B,
       author = {{Barvainis}, Richard},
        title = "{Hot Dust and the Near-Infrared Bump in the Continuum Spectra of Quasars and Active Galactic Nuclei}",
      journal = {\apj},
         year = 1987,
        month = sep,
       volume = {320},
        pages = {537},
          doi = {10.1086/165571},
       adsurl = {https://ui.adsabs.harvard.edu/abs/1987ApJ...320..537B}
}

@ARTICLE{1993ApJ...412..254C,
       author = {{Chen}, Xingming and {Taam}, Ronald E.},
        title = "{The Structure and Stability of Transonic Accretion Disks Surrounding Black Holes}",
      journal = {\apj},
         year = 1993,
        month = jul,
       volume = {412},
        pages = {254},
          doi = {10.1086/172916},
       adsurl = {https://ui.adsabs.harvard.edu/abs/1993ApJ...412..254C}
}

@ARTICLE{1992ARA&A..30..575C,
       author = {{Condon}, J.~J.},
        title = "{Radio emission from normal galaxies.}",
      journal = {\araa},
         year = 1992,
        month = jan,
       volume = {30},
        pages = {575-611},
          doi = {10.1146/annurev.aa.30.090192.003043},
       adsurl = {https://ui.adsabs.harvard.edu/abs/1992ARA&A..30..575C}
}

@ARTICLE{2011ApJ...727L..24D,
       author = {{Dexter}, Jason and {Agol}, Eric},
        title = "{Quasar Accretion Disks are Strongly Inhomogeneous}",
      journal = {\apjl},
         year = 2011,
        month = jan,
       volume = {727},
       number = {1},
          eid = {L24},
        pages = {L24},
          doi = {10.1088/2041-8205/727/1/L24},
archivePrefix = {arXiv},
       eprint = {1012.3169},
 primaryClass = {astro-ph.CO},
       adsurl = {https://ui.adsabs.harvard.edu/abs/2011ApJ...727L..24D}
}

@ARTICLE{2014arXiv1412.4872D,
       author = {{Dor{\'e}}, Olivier and {Bock}, Jamie and {Ashby}, Matthew and {Capak}, Peter and {Cooray}, Asantha and {de Putter}, Roland and {Eifler}, Tim and {Flagey}, Nicolas and {Gong}, Yan and {Habib}, Salman and {Heitmann}, Katrin and {Hirata}, Chris and {Jeong}, Woong-Seob and {Katti}, Raj and {Korngut}, Phil and {Krause}, Elisabeth and {Lee}, Dae-Hee and {Masters}, Daniel and {Mauskopf}, Phil and {Melnick}, Gary and {Mennesson}, Bertrand and {Nguyen}, Hien and {{\"O}berg}, Karin and {Pullen}, Anthony and {Raccanelli}, Alvise and {Smith}, Roger and {Song}, Yong-Seon and {Tolls}, Volker and {Unwin}, Steve and {Venumadhav}, Tejaswi and {Viero}, Marco and {Werner}, Mike and {Zemcov}, Mike},
        title = "{Cosmology with the SPHEREX All-Sky Spectral Survey}",
      journal = {arXiv e-prints},
         year = 2014,
        month = dec,
          eid = {arXiv:1412.4872},
        pages = {arXiv:1412.4872},
          doi = {10.48550/arXiv.1412.4872},
archivePrefix = {arXiv},
       eprint = {1412.4872},
 primaryClass = {astro-ph.CO},
       adsurl = {https://ui.adsabs.harvard.edu/abs/2014arXiv1412.4872D}
}

@ARTICLE{2016ApJ...832..188D,
       author = {{Dou}, Liming and {Wang}, Ting-gui and {Jiang}, Ning and {Yang}, Chenwei and {Lyu}, Jianwei and {Zhou}, Hongyan},
        title = "{Long Fading Mid-infrared Emission in Transient Coronal Line Emitters: Dust Echo of a Tidal Disruption Flare}",
      journal = {\apj},
         year = 2016,
        month = dec,
       volume = {832},
       number = {2},
          eid = {188},
        pages = {188},
          doi = {10.3847/0004-637X/832/2/188},
archivePrefix = {arXiv},
       eprint = {1605.05145},
 primaryClass = {astro-ph.HE},
       adsurl = {https://ui.adsabs.harvard.edu/abs/2016ApJ...832..188D}
}

@ARTICLE{2022A&A...662A.112E,
       author = {{Euclid Collaboration} and {Scaramella}, R. and {Amiaux}, J. and {Mellier}, Y. and {Burigana}, C. and {Carvalho}, C.~S. and {Cuillandre}, J.-C. and {Da Silva}, A. and {Derosa}, A. and {Dinis}, J. and {Maiorano}, E. and {Maris}, M. and {Tereno}, I. and {Laureijs}, R. and {Boenke}, T. and {Buenadicha}, G. and {Dupac}, X. and {Gaspar Venancio}, L.~M. and {G{\'o}mez-{\'A}lvarez}, P. and {Hoar}, J. and {Lorenzo Alvarez}, J. and {Racca}, G.~D. and {Saavedra-Criado}, G. and {Schwartz}, J. and {Vavrek}, R. and {Schirmer}, M. and {Aussel}, H. and {Azzollini}, R. and {Cardone}, V.~F. and {Cropper}, M. and {Ealet}, A. and {Garilli}, B. and {Gillard}, W. and {Granett}, B.~R. and {Guzzo}, L. and {Hoekstra}, H. and {Jahnke}, K. and {Kitching}, T. and {Maciaszek}, T. and {Meneghetti}, M. and {Miller}, L. and {Nakajima}, R. and {Niemi}, S.~M. and {Pasian}, F. and {Percival}, W.~J. and {Pottinger}, S. and {Sauvage}, M. and {Scodeggio}, M. and {Wachter}, S. and {Zacchei}, A. and {Aghanim}, N. and {Amara}, A. and {Auphan}, T. and {Auricchio}, N. and {Awan}, S. and {Balestra}, A. and {Bender}, R. and {Bodendorf}, C. and {Bonino}, D. and {Branchini}, E. and {Brau-Nogue}, S. and {Brescia}, M. and {Candini}, G.~P. and {Capobianco}, V. and {Carbone}, C. and {Carlberg}, R.~G. and {Carretero}, J. and {Casas}, R. and {Castander}, F.~J. and {Castellano}, M. and {Cavuoti}, S. and {Cimatti}, A. and {Cledassou}, R. and {Congedo}, G. and {Conselice}, C.~J. and {Conversi}, L. and {Copin}, Y. and {Corcione}, L. and {Costille}, A. and {Courbin}, F. and {Degaudenzi}, H. and {Douspis}, M. and {Dubath}, F. and {Duncan}, C.~A.~J. and {Dusini}, S. and {Farrens}, S. and {Ferriol}, S. and {Fosalba}, P. and {Fourmanoit}, N. and {Frailis}, M. and {Franceschi}, E. and {Franzetti}, P. and {Fumana}, M. and {Gillis}, B. and {Giocoli}, C. and {Grazian}, A. and {Grupp}, F. and {Haugan}, S.~V.~H. and {Holmes}, W. and {Hormuth}, F. and {Hudelot}, P. and {Kermiche}, S. and {Kiessling}, A. and {Kilbinger}, M. and {Kohley}, R. and {Kubik}, B. and {K{\"u}mmel}, M. and {Kunz}, M. and {Kurki-Suonio}, H. and {Lahav}, O. and {Ligori}, S. and {Lilje}, P.~B. and {Lloro}, I. and {Mansutti}, O. and {Marggraf}, O. and {Markovic}, K. and {Marulli}, F. and {Massey}, R. and {Maurogordato}, S. and {Melchior}, M. and {Merlin}, E. and {Meylan}, G. and {Mohr}, J.~J. and {Moresco}, M. and {Morin}, B. and {Moscardini}, L. and {Munari}, E. and {Nichol}, R.~C. and {Padilla}, C. and {Paltani}, S. and {Peacock}, J. and {Pedersen}, K. and {Pettorino}, V. and {Pires}, S. and {Poncet}, M. and {Popa}, L. and {Pozzetti}, L. and {Raison}, F. and {Rebolo}, R. and {Rhodes}, J. and {Rix}, H.-W. and {Roncarelli}, M. and {Rossetti}, E. and {Saglia}, R. and {Schneider}, P. and {Schrabback}, T. and {Secroun}, A. and {Seidel}, G. and {Serrano}, S. and {Sirignano}, C. and {Sirri}, G. and {Skottfelt}, J. and {Stanco}, L. and {Starck}, J.~L. and {Tallada-Cresp{\'\i}}, P. and {Tavagnacco}, D. and {Taylor}, A.~N. and {Teplitz}, H.~I. and {Toledo-Moreo}, R. and {Torradeflot}, F. and {Trifoglio}, M. and {Valentijn}, E.~A. and {Valenziano}, L. and {Verdoes Kleijn}, G.~A. and {Wang}, Y. and {Welikala}, N. and {Weller}, J. and {Wetzstein}, M. and {Zamorani}, G. and {Zoubian}, J. and {Andreon}, S. and {Baldi}, M. and {Bardelli}, S. and {Boucaud}, A. and {Camera}, S. and {Di Ferdinando}, D. and {Fabbian}, G. and {Farinelli}, R. and {Galeotta}, S. and {Graci{\'a}-Carpio}, J. and {Maino}, D. and {Medinaceli}, E. and {Mei}, S. and {Neissner}, C. and {Polenta}, G. and {Renzi}, A. and {Romelli}, E. and {Rosset}, C. and {Sureau}, F. and {Tenti}, M. and {Vassallo}, T. and {Zucca}, E. and {Baccigalupi}, C. and {Balaguera-Antol{\'\i}nez}, A. and {Battaglia}, P. and {Biviano}, A. and {Borgani}, S. and {Bozzo}, E. and {Cabanac}, R. and {Cappi}, A.},
        title = "{Euclid preparation. I. The Euclid Wide Survey}",
      journal = {\aap},
         year = 2022,
        month = jun,
       volume = {662},
          eid = {A112},
        pages = {A112},
          doi = {10.1051/0004-6361/202141938},
archivePrefix = {arXiv},
       eprint = {2108.01201},
 primaryClass = {astro-ph.CO},
       adsurl = {https://ui.adsabs.harvard.edu/abs/2022A&A...662A.112E}
}

@ARTICLE{2010ApJS..189...37E,
       author = {{Evans}, Ian N. and {Primini}, Francis A. and {Glotfelty}, Kenny J. and {Anderson}, Craig S. and {Bonaventura}, Nina R. and {Chen}, Judy C. and {Davis}, John E. and {Doe}, Stephen M. and {Evans}, Janet D. and {Fabbiano}, Giuseppina and {Galle}, Elizabeth C. and {Gibbs}, II, Danny G. and {Grier}, John D. and {Hain}, Roger M. and {Hall}, Diane M. and {Harbo}, Peter N. and {He}, Xiangqun Helen and {Houck}, John C. and {Karovska}, Margarita and {Kashyap}, Vinay L. and {Lauer}, Jennifer and {McCollough}, Michael L. and {McDowell}, Jonathan C. and {Miller}, Joseph B. and {Mitschang}, Arik W. and {Morgan}, Douglas L. and {Mossman}, Amy E. and {Nichols}, Joy S. and {Nowak}, Michael A. and {Plummer}, David A. and {Refsdal}, Brian L. and {Rots}, Arnold H. and {Siemiginowska}, Aneta and {Sundheim}, Beth A. and {Tibbetts}, Michael S. and {Van Stone}, David W. and {Winkelman}, Sherry L. and {Zografou}, Panagoula},
        title = "{The Chandra Source Catalog}",
      journal = {\apjs},
         year = 2010,
        month = jul,
       volume = {189},
       number = {1},
        pages = {37-82},
          doi = {10.1088/0067-0049/189/1/37},
archivePrefix = {arXiv},
       eprint = {1005.4665},
 primaryClass = {astro-ph.HE},
       adsurl = {https://ui.adsabs.harvard.edu/abs/2010ApJS..189...37E}
}

@ARTICLE{2023MNRAS.518..174E,
       author = {{Evans}, P.~A. and {Page}, K.~L. and {Beardmore}, A.~P. and {Eyles-Ferris}, R.~A.~J. and {Osborne}, J.~P. and {Campana}, S. and {Kennea}, J.~A. and {Cenko}, S.~B.},
        title = "{A real-time transient detector and the living Swift-XRT point source catalogue}",
      journal = {\mnras},
         year = 2023,
        month = jan,
       volume = {518},
       number = {1},
        pages = {174-184},
          doi = {10.1093/mnras/stac2937},
archivePrefix = {arXiv},
       eprint = {2208.14478},
 primaryClass = {astro-ph.HE},
       adsurl = {https://ui.adsabs.harvard.edu/abs/2023MNRAS.518..174E}
}

@ARTICLE{1992MNRAS.256P..23G,
       author = {{Glass}, I.~S.},
        title = "{Infrared variability of the Seyfert galaxy NGC 3783.}",
      journal = {\mnras},
         year = 1992,
        month = may,
       volume = {256},
        pages = {23P-27},
          doi = {10.1093/mnras/256.1.23P},
       adsurl = {https://ui.adsabs.harvard.edu/abs/1992MNRAS.256P..23G}
}

@ARTICLE{Matplotlib,  
	author={J. D. {Hunter}},  
	journal={Computing in Science   Engineering},   
	title={Matplotlib: A 2D Graphics Environment},   
	year={2007},  
	volume={9},  
	number={3},  
	pages={90-95}
}

@ARTICLE{2019ApJ...873..111I,
       author = {{Ivezi{\'c}}, {\v{Z}}eljko and {Kahn}, Steven M. and {Tyson}, J. Anthony and {Abel}, Bob and {Acosta}, Emily and {Allsman}, Robyn and {Alonso}, David and {AlSayyad}, Yusra and {Anderson}, Scott F. and {Andrew}, John and {Angel}, James Roger P. and {Angeli}, George Z. and {Ansari}, Reza and {Antilogus}, Pierre and {Araujo}, Constanza and {Armstrong}, Robert and {Arndt}, Kirk T. and {Astier}, Pierre and {Aubourg}, {\'E}ric and {Auza}, Nicole and {Axelrod}, Tim S. and {Bard}, Deborah J. and {Barr}, Jeff D. and {Barrau}, Aurelian and {Bartlett}, James G. and {Bauer}, Amanda E. and {Bauman}, Brian J. and {Baumont}, Sylvain and {Bechtol}, Ellen and {Bechtol}, Keith and {Becker}, Andrew C. and {Becla}, Jacek and {Beldica}, Cristina and {Bellavia}, Steve and {Bianco}, Federica B. and {Biswas}, Rahul and {Blanc}, Guillaume and {Blazek}, Jonathan and {Blandford}, Roger D. and {Bloom}, Josh S. and {Bogart}, Joanne and {Bond}, Tim W. and {Booth}, Michael T. and {Borgland}, Anders W. and {Borne}, Kirk and {Bosch}, James F. and {Boutigny}, Dominique and {Brackett}, Craig A. and {Bradshaw}, Andrew and {Brandt}, William Nielsen and {Brown}, Michael E. and {Bullock}, James S. and {Burchat}, Patricia and {Burke}, David L. and {Cagnoli}, Gianpietro and {Calabrese}, Daniel and {Callahan}, Shawn and {Callen}, Alice L. and {Carlin}, Jeffrey L. and {Carlson}, Erin L. and {Chandrasekharan}, Srinivasan and {Charles-Emerson}, Glenaver and {Chesley}, Steve and {Cheu}, Elliott C. and {Chiang}, Hsin-Fang and {Chiang}, James and {Chirino}, Carol and {Chow}, Derek and {Ciardi}, David R. and {Claver}, Charles F. and {Cohen-Tanugi}, Johann and {Cockrum}, Joseph J. and {Coles}, Rebecca and {Connolly}, Andrew J. and {Cook}, Kem H. and {Cooray}, Asantha and {Covey}, Kevin R. and {Cribbs}, Chris and {Cui}, Wei and {Cutri}, Roc and {Daly}, Philip N. and {Daniel}, Scott F. and {Daruich}, Felipe and {Daubard}, Guillaume and {Daues}, Greg and {Dawson}, William and {Delgado}, Francisco and {Dellapenna}, Alfred and {de Peyster}, Robert and {de Val-Borro}, Miguel and {Digel}, Seth W. and {Doherty}, Peter and {Dubois}, Richard and {Dubois-Felsmann}, Gregory P. and {Durech}, Josef and {Economou}, Frossie and {Eifler}, Tim and {Eracleous}, Michael and {Emmons}, Benjamin L. and {Fausti Neto}, Angelo and {Ferguson}, Henry and {Figueroa}, Enrique and {Fisher-Levine}, Merlin and {Focke}, Warren and {Foss}, Michael D. and {Frank}, James and {Freemon}, Michael D. and {Gangler}, Emmanuel and {Gawiser}, Eric and {Geary}, John C. and {Gee}, Perry and {Geha}, Marla and {Gessner}, Charles J.~B. and {Gibson}, Robert R. and {Gilmore}, D. Kirk and {Glanzman}, Thomas and {Glick}, William and {Goldina}, Tatiana and {Goldstein}, Daniel A. and {Goodenow}, Iain and {Graham}, Melissa L. and {Gressler}, William J. and {Gris}, Philippe and {Guy}, Leanne P. and {Guyonnet}, Augustin and {Haller}, Gunther and {Harris}, Ron and {Hascall}, Patrick A. and {Haupt}, Justine and {Hernandez}, Fabio and {Herrmann}, Sven and {Hileman}, Edward and {Hoblitt}, Joshua and {Hodgson}, John A. and {Hogan}, Craig and {Howard}, James D. and {Huang}, Dajun and {Huffer}, Michael E. and {Ingraham}, Patrick and {Innes}, Walter R. and {Jacoby}, Suzanne H. and {Jain}, Bhuvnesh and {Jammes}, Fabrice and {Jee}, M. James and {Jenness}, Tim and {Jernigan}, Garrett and {Jevremovi{\'c}}, Darko and {Johns}, Kenneth and {Johnson}, Anthony S. and {Johnson}, Margaret W.~G. and {Jones}, R. Lynne and {Juramy-Gilles}, Claire and {Juri{\'c}}, Mario and {Kalirai}, Jason S. and {Kallivayalil}, Nitya J. and {Kalmbach}, Bryce and {Kantor}, Jeffrey P. and {Karst}, Pierre and {Kasliwal}, Mansi M. and {Kelly}, Heather and {Kessler}, Richard and {Kinnison}, Veronica and {Kirkby}, David and {Knox}, Lloyd and {Kotov}, Ivan V. and {Krabbendam}, Victor L. and {Krughoff}, K. Simon and {Kub{\'a}nek}, Petr and {Kuczewski}, John and {Kulkarni}, Shri and {Ku}, John and {Kurita}, Nadine R. and {Lage}, Craig S. and {Lambert}, Ron and {Lange}, Travis and {Langton}, J. Brian and {Le Guillou}, Laurent and {Levine}, Deborah and {Liang}, Ming and {Lim}, Kian-Tat and {Lintott}, Chris J. and {Long}, Kevin E. and {Lopez}, Margaux and {Lotz}, Paul J. and {Lupton}, Robert H. and {Lust}, Nate B. and {MacArthur}, Lauren A. and {Mahabal}, Ashish and {Mandelbaum}, Rachel and {Markiewicz}, Thomas W. and {Marsh}, Darren S. and {Marshall}, Philip J. and {Marshall}, Stuart and {May}, Morgan and {McKercher}, Robert and {McQueen}, Michelle and {Meyers}, Joshua and {Migliore}, Myriam and {Miller}, Michelle and {Mills}, David J.},
        title = "{LSST: From Science Drivers to Reference Design and Anticipated Data Products}",
      journal = {\apj},
         year = 2019,
        month = mar,
       volume = {873},
       number = {2},
          eid = {111},
        pages = {111},
          doi = {10.3847/1538-4357/ab042c},
archivePrefix = {arXiv},
       eprint = {0805.2366},
 primaryClass = {astro-ph},
       adsurl = {https://ui.adsabs.harvard.edu/abs/2019ApJ...873..111I}
}

@ARTICLE{2021ApJS..252...32J,
       author = {{Jiang}, Ning and {Wang}, Tinggui and {Dou}, Liming and {Shu}, Xinwen and {Hu}, Xueyang and {Liu}, Hui and {Wang}, Yibo and {Yan}, Lin and {Sheng}, Zhenfeng and {Yang}, Chenwei and {Sun}, Luming and {Zhou}, Hongyan},
        title = "{Mid-infrared Outbursts in Nearby Galaxies (MIRONG). I. Sample Selection and Characterization}",
      journal = {\apjs},
         year = 2021,
        month = feb,
       volume = {252},
       number = {2},
          eid = {32},
        pages = {32},
          doi = {10.3847/1538-4365/abd1dc},
archivePrefix = {arXiv},
       eprint = {2012.06806},
 primaryClass = {astro-ph.GA},
       adsurl = {https://ui.adsabs.harvard.edu/abs/2021ApJS..252...32J}
}

@ARTICLE{2011ApJ...737..105K,
       author = {{Kawaguchi}, Toshihiro and {Mori}, Masao},
        title = "{Near-infrared Reverberation by Dusty Clumpy Tori in Active Galactic Nuclei}",
      journal = {\apj},
         year = 2011,
        month = aug,
       volume = {737},
       number = {2},
          eid = {105},
        pages = {105},
          doi = {10.1088/0004-637X/737/2/105},
archivePrefix = {arXiv},
       eprint = {1107.0678},
 primaryClass = {astro-ph.CO},
       adsurl = {https://ui.adsabs.harvard.edu/abs/2011ApJ...737..105K}
}

@ARTICLE{2009ApJ...698..895K,
       author = {{Kelly}, Brandon C. and {Bechtold}, Jill and {Siemiginowska}, Aneta},
        title = "{Are the Variations in Quasar Optical Flux Driven by Thermal Fluctuations?}",
      journal = {\apj},
         year = 2009,
        month = jun,
       volume = {698},
       number = {1},
        pages = {895-910},
          doi = {10.1088/0004-637X/698/1/895},
archivePrefix = {arXiv},
       eprint = {0903.5315},
 primaryClass = {astro-ph.CO},
       adsurl = {https://ui.adsabs.harvard.edu/abs/2009ApJ...698..895K}
}

@ARTICLE{2026ApJ..1000..245K,
       author = {{Kim}, Shinyu and {Kim}, Minjin and {Son}, Suyeon and {Ho}, Luis C.},
        title = "{Mid-infrared Variability-based Active Galactic Nucleus Selection Using the Multiepoch Photometric Data from WISE}",
      journal = {\apj},
         year = 2026,
        month = apr,
       volume = {1000},
       number = {2},
          eid = {245},
        pages = {245},
          doi = {10.3847/1538-4357/ae4ecd},
archivePrefix = {arXiv},
       eprint = {2603.06227},
 primaryClass = {astro-ph.GA},
       adsurl = {https://ui.adsabs.harvard.edu/abs/2026ApJ..1000..245K}
}

@ARTICLE{2014ApJ...788..159K,
       author = {{Koshida}, Shintaro and {Minezaki}, Takeo and {Yoshii}, Yuzuru and {Kobayashi}, Yukiyasu and {Sakata}, Yu and {Sugawara}, Shota and {Enya}, Keigo and {Suganuma}, Masahiro and {Tomita}, Hiroyuki and {Aoki}, Tsutomu and {Peterson}, Bruce A.},
        title = "{Reverberation Measurements of the Inner Radius of the Dust Torus in 17 Seyfert Galaxies}",
      journal = {\apj},
         year = 2014,
        month = jun,
       volume = {788},
       number = {2},
          eid = {159},
        pages = {159},
          doi = {10.1088/0004-637X/788/2/159},
archivePrefix = {arXiv},
       eprint = {1406.2078},
 primaryClass = {astro-ph.GA},
       adsurl = {https://ui.adsabs.harvard.edu/abs/2014ApJ...788..159K}
}

@ARTICLE{2010ApJ...716..530K,
       author = {{Koz{\l}owski}, Szymon and {Kochanek}, Christopher S. and {Stern}, Daniel and {Ashby}, Matthew L.~N. and {Assef}, Roberto J. and {Bock}, J.~J. and {Borys}, C. and {Brand}, K. and {Brodwin}, M. and {Brown}, M.~J.~I. and {Cool}, R. and {Cooray}, A. and {Croft}, S. and {Dey}, Arjun and {Eisenhardt}, P.~R. and {Gonzalez}, A. and {Gorjian}, V. and {Griffith}, R. and {Grogin}, N. and {Ivison}, R. and {Jacob}, J. and {Jannuzi}, B.~T. and {Mainzer}, A. and {Moustakas}, L. and {R{\"o}ttgering}, H. and {Seymour}, N. and {Smith}, H.~A. and {Stanford}, S.~A. and {Stauffer}, J.~R. and {Sullivan}, I.~S. and {van Breugel}, W. and {Willner}, S.~P. and {Wright}, E.~L.},
        title = "{Mid-infrared Variability from the Spitzer Deep Wide-field Survey}",
      journal = {\apj},
         year = 2010,
        month = jun,
       volume = {716},
       number = {1},
        pages = {530-543},
          doi = {10.1088/0004-637X/716/1/530},
archivePrefix = {arXiv},
       eprint = {1002.3365},
 primaryClass = {astro-ph.CO},
       adsurl = {https://ui.adsabs.harvard.edu/abs/2010ApJ...716..530K}
}

@ARTICLE{2016ApJ...817..119K,
       author = {{Koz{\l}owski}, Szymon and {Kochanek}, Christopher S. and {Ashby}, Matthew L.~N. and {Assef}, Roberto J. and {Brodwin}, Mark and {Eisenhardt}, Peter R. and {Jannuzi}, Buell T. and {Stern}, Daniel},
        title = "{Quasar Variability in the Mid-Infrared}",
      journal = {\apj},
         year = 2016,
        month = feb,
       volume = {817},
       number = {2},
          eid = {119},
        pages = {119},
          doi = {10.3847/0004-637X/817/2/119},
archivePrefix = {arXiv},
       eprint = {1509.02467},
 primaryClass = {astro-ph.GA},
       adsurl = {https://ui.adsabs.harvard.edu/abs/2016ApJ...817..119K}
}

@ARTICLE{2016ApJ...826..118K,
       author = {{Koz{\l}owski}, Szymon},
        title = "{Revisiting Stochastic Variability of AGNs with Structure Functions}",
      journal = {\apj},
         year = 2016,
        month = aug,
       volume = {826},
       number = {2},
          eid = {118},
        pages = {118},
          doi = {10.3847/0004-637X/826/2/118},
archivePrefix = {arXiv},
       eprint = {1604.05858},
 primaryClass = {astro-ph.GA},
       adsurl = {https://ui.adsabs.harvard.edu/abs/2016ApJ...826..118K}
}

@ARTICLE{2015ApJ...800..144L,
       author = {{LaMassa}, Stephanie M. and {Cales}, Sabrina and {Moran}, Edward C. and {Myers}, Adam D. and {Richards}, Gordon T. and {Eracleous}, Michael and {Heckman}, Timothy M. and {Gallo}, Luigi and {Urry}, C. Megan},
        title = "{The Discovery of the First {\textquotedblleft}Changing Look{\textquotedblright} Quasar: New Insights Into the Physics and Phenomenology of Active Galactic Nucleus}",
      journal = {\apj},
         year = 2015,
        month = feb,
       volume = {800},
       number = {2},
          eid = {144},
        pages = {144},
          doi = {10.1088/0004-637X/800/2/144},
archivePrefix = {arXiv},
       eprint = {1412.2136},
 primaryClass = {astro-ph.GA},
       adsurl = {https://ui.adsabs.harvard.edu/abs/2015ApJ...800..144L}
}

@misc{leung2026discoveryquasarvariabilityearly,
      title={Discovery of Quasar Variability and Early Accretion Disk Signatures at Cosmic Dawn}, 
      author={Gene C. K. Leung and Anna-Christina Eilers and Christos Panagiotou and Julien Wolf and Kishalay De and Luke Weisenbach and Minghao Yue and Xiaohui Fan and Yuzo Ishikawa and Erin Kara and Mirko Krumpe and Andrea Merloni and Robert A. Simcoe and Feige Wang and Jinyi Yang},
      year={2026},
      eprint={2605.00978},
      archivePrefix={arXiv},
      primaryClass={astro-ph.GA},
      url={https://arxiv.org/abs/2605.00978}, 
}

@ARTICLE{2022A&A...661A...5L,
       author = {{Liu}, Teng and {Buchner}, Johannes and {Nandra}, Kirpal and {Merloni}, Andrea and {Dwelly}, Tom and {Sanders}, Jeremy S. and {Salvato}, Mara and {Arcodia}, Riccardo and {Brusa}, Marcella and {Wolf}, Julien and {Georgakakis}, Antonis and {Boller}, Thomas and {Krumpe}, Mirko and {Lamer}, Georg and {Waddell}, Sophia and {Urrutia}, Tanya and {Schwope}, Axel and {Robrade}, Jan and {Wilms}, J{\"o}rn and {Dauser}, Thomas and {Comparat}, Johan and {Toba}, Yoshiki and {Ichikawa}, Kohei and {Iwasawa}, Kazushi and {Shen}, Yue and {Medel}, Hector Ibarra},
        title = "{The eROSITA Final Equatorial-Depth Survey (eFEDS). The AGN catalog and its X-ray spectral properties}",
      journal = {\aap},
         year = 2022,
        month = may,
       volume = {661},
          eid = {A5},
        pages = {A5},
          doi = {10.1051/0004-6361/202141643},
archivePrefix = {arXiv},
       eprint = {2106.14522},
 primaryClass = {astro-ph.HE},
       adsurl = {https://ui.adsabs.harvard.edu/abs/2022A&A...661A...5L}
}

@ARTICLE{1969Natur.223..690L,
       author = {{Lynden-Bell}, D.},
        title = "{Galactic Nuclei as Collapsed Old Quasars}",
      journal = {\nat},
         year = 1969,
        month = aug,
       volume = {223},
       number = {5207},
        pages = {690-694},
          doi = {10.1038/223690a0},
       adsurl = {https://ui.adsabs.harvard.edu/abs/1969Natur.223..690L}
}

@ARTICLE{2019ApJ...886...33L,
       author = {{Lyu}, Jianwei and {Rieke}, George H. and {Smith}, Paul S.},
        title = "{Mid-IR Variability and Dust Reverberation Mapping of Low-z Quasars. I. Data, Methods, and Basic Results}",
      journal = {\apj},
         year = 2019,
        month = nov,
       volume = {886},
       number = {1},
          eid = {33},
        pages = {33},
          doi = {10.3847/1538-4357/ab481d},
archivePrefix = {arXiv},
       eprint = {1909.11101},
 primaryClass = {astro-ph.GA},
       adsurl = {https://ui.adsabs.harvard.edu/abs/2019ApJ...886...33L}
}

@ARTICLE{2019ApJ...874....8M,
       author = {{MacLeod}, Chelsea L. and {Green}, Paul J. and {Anderson}, Scott F. and {Bruce}, Alastair and {Eracleous}, Michael and {Graham}, Matthew and {Homan}, David and {Lawrence}, Andy and {LeBleu}, Amy and {Ross}, Nicholas P. and {Ruan}, John J. and {Runnoe}, Jessie and {Stern}, Daniel and {Burgett}, William and {Chambers}, Kenneth C. and {Kaiser}, Nick and {Magnier}, Eugene and {Metcalfe}, Nigel},
        title = "{Changing-look Quasar Candidates: First Results from Follow-up Spectroscopy of Highly Optically Variable Quasars}",
      journal = {\apj},
         year = 2019,
        month = mar,
       volume = {874},
       number = {1},
          eid = {8},
        pages = {8},
          doi = {10.3847/1538-4357/ab05e2},
archivePrefix = {arXiv},
       eprint = {1810.00087},
 primaryClass = {astro-ph.GA},
       adsurl = {https://ui.adsabs.harvard.edu/abs/2019ApJ...874....8M}
}

@ARTICLE{2010ApJ...721.1014M,
       author = {{MacLeod}, C.~L. and {Ivezi{\'c}}, {\v{Z}}. and {Kochanek}, C.~S. and {Koz{\l}owski}, S. and {Kelly}, B. and {Bullock}, E. and {Kimball}, A. and {Sesar}, B. and {Westman}, D. and {Brooks}, K. and {Gibson}, R. and {Becker}, A.~C. and {de Vries}, W.~H.},
        title = "{Modeling the Time Variability of SDSS Stripe 82 Quasars as a Damped Random Walk}",
      journal = {\apj},
         year = 2010,
        month = oct,
       volume = {721},
       number = {2},
        pages = {1014-1033},
          doi = {10.1088/0004-637X/721/2/1014},
archivePrefix = {arXiv},
       eprint = {1004.0276},
 primaryClass = {astro-ph.CO},
       adsurl = {https://ui.adsabs.harvard.edu/abs/2010ApJ...721.1014M}
}

@ARTICLE{1963ApJ...138...30M,
       author = {{Matthews}, Thomas A. and {Sandage}, Allan R.},
        title = "{Optical Identification of 3C 48, 3C 196, and 3C 286 with Stellar Objects.}",
      journal = {\apj},
         year = 1963,
        month = jul,
       volume = {138},
        pages = {30},
          doi = {10.1086/147615},
       adsurl = {https://ui.adsabs.harvard.edu/abs/1963ApJ...138...30M}
}

@ARTICLE{1996ApJ...473..763M,
       author = {{McLaughlin}, M.~A. and {Mattox}, J.~R. and {Cordes}, J.~M. and {Thompson}, D.~J.},
        title = "{Variability of CGRO/EGRET Gamma-Ray Sources}",
      journal = {\apj},
         year = 1996,
        month = dec,
       volume = {473},
        pages = {763},
          doi = {10.1086/178188},
       adsurl = {https://ui.adsabs.harvard.edu/abs/1996ApJ...473..763M}
}

@ARTICLE{2018Sci...361..482M,
       author = {{Mattila}, S. and {P{\'e}rez-Torres}, M. and {Efstathiou}, A. and {Mimica}, P. and {Fraser}, M. and {Kankare}, E. and {Alberdi}, A. and {Aloy}, M. {\'A}. and {Heikkil{\"a}}, T. and {Jonker}, P.~G. and {Lundqvist}, P. and {Mart{\'\i}-Vidal}, I. and {Meikle}, W.~P.~S. and {Romero-Ca{\~n}izales}, C. and {Smartt}, S.~J. and {Tsygankov}, S. and {Varenius}, E. and {Alonso-Herrero}, A. and {Bondi}, M. and {Fransson}, C. and {Herrero-Illana}, R. and {Kangas}, T. and {Kotak}, R. and {Ram{\'\i}rez-Olivencia}, N. and {V{\"a}is{\"a}nen}, P. and {Beswick}, R.~J. and {Clements}, D.~L. and {Greimel}, R. and {Harmanen}, J. and {Kotilainen}, J. and {Nandra}, K. and {Reynolds}, T. and {Ryder}, S. and {Walton}, N.~A. and {Wiik}, K. and {{\"O}stlin}, G.},
        title = "{A dust-enshrouded tidal disruption event with a resolved radio jet in a galaxy merger}",
      journal = {Science},
         year = 2018,
        month = aug,
       volume = {361},
       number = {6401},
        pages = {482-485},
          doi = {10.1126/science.aao4669},
archivePrefix = {arXiv},
       eprint = {1806.05717},
 primaryClass = {astro-ph.GA},
       adsurl = {https://ui.adsabs.harvard.edu/abs/2018Sci...361..482M}
}

@ARTICLE{2025A&A...695A.228N,
       author = {{Necker}, J. and {Graikou}, E. and {Kowalski}, M. and {Franckowiak}, A. and {Nordin}, J. and {Pernice}, T. and {van Velzen}, S. and {Veres}, P.~M.},
        title = "{Flaires: A comprehensive catalog of dust echo-like infrared flares}",
      journal = {\aap},
         year = 2025,
        month = mar,
       volume = {695},
          eid = {A228},
        pages = {A228},
          doi = {10.1051/0004-6361/202451340},
archivePrefix = {arXiv},
       eprint = {2407.01039},
 primaryClass = {astro-ph.HE},
       adsurl = {https://ui.adsabs.harvard.edu/abs/2025A&A...695A.228N}
}

@ARTICLE{2015ARA&A..53..365N,
       author = {{Netzer}, Hagai},
        title = "{Revisiting the Unified Model of Active Galactic Nuclei}",
      journal = {\araa},
         year = 2015,
        month = aug,
       volume = {53},
        pages = {365-408},
          doi = {10.1146/annurev-astro-082214-122302},
archivePrefix = {arXiv},
       eprint = {1505.00811},
 primaryClass = {astro-ph.GA},
       adsurl = {https://ui.adsabs.harvard.edu/abs/2015ARA&A..53..365N}
}

@ARTICLE{1999AJ....118...35N,
       author = {{Neugebauer}, G. and {Matthews}, K.},
        title = "{Variability of Quasars at 10 Microns}",
      journal = {\aj},
         year = 1999,
        month = jul,
       volume = {118},
       number = {1},
        pages = {35-45},
          doi = {10.1086/300945},
archivePrefix = {arXiv},
       eprint = {astro-ph/9903363},
 primaryClass = {astro-ph},
       adsurl = {https://ui.adsabs.harvard.edu/abs/1999AJ....118...35N}
}

@ARTICLE{2024ApJ...977..102P,
       author = {{Pai}, Aashay and {Blanton}, Michael R. and {Moustakas}, John},
        title = "{Mid-infrared Variability in Nearby Galaxies from the MaNGA Sample}",
      journal = {\apj},
         year = 2024,
        month = dec,
       volume = {977},
       number = {1},
          eid = {102},
        pages = {102},
          doi = {10.3847/1538-4357/ad89b8},
archivePrefix = {arXiv},
       eprint = {2407.05508},
 primaryClass = {astro-ph.GA},
       adsurl = {https://ui.adsabs.harvard.edu/abs/2024ApJ...977..102P}
}

@ARTICLE{2017ApJ...842...96R,
       author = {{Rakshit}, Suvendu and {Stalin}, C.~S.},
        title = "{Optical Variability of Narrow-line and Broad-line Seyfert 1 Galaxies}",
      journal = {\apj},
         year = 2017,
        month = jun,
       volume = {842},
       number = {2},
          eid = {96},
        pages = {96},
          doi = {10.3847/1538-4357/aa72f4},
archivePrefix = {arXiv},
       eprint = {1705.05123},
 primaryClass = {astro-ph.GA},
       adsurl = {https://ui.adsabs.harvard.edu/abs/2017ApJ...842...96R}
}

@ARTICLE{2019MNRAS.483.2362R,
       author = {{Rakshit}, Suvendu and {Johnson}, Ansu and {Stalin}, C.~S. and {Gandhi}, Poshak and {Hoenig}, Sebastian},
        title = "{WISE view of narrow-line Seyfert 1 galaxies: mid-infrared colour and variability}",
      journal = {\mnras},
         year = 2019,
        month = feb,
       volume = {483},
       number = {2},
        pages = {2362-2370},
          doi = {10.1093/mnras/sty3261},
archivePrefix = {arXiv},
       eprint = {1811.11372},
 primaryClass = {astro-ph.GA},
       adsurl = {https://ui.adsabs.harvard.edu/abs/2019MNRAS.483.2362R}
}

@ARTICLE{2020ApJS..249...17R,
       author = {{Rakshit}, Suvendu and {Stalin}, C.~S. and {Kotilainen}, Jari},
        title = "{Spectral Properties of Quasars from Sloan Digital Sky Survey Data Release 14: The Catalog}",
      journal = {\apjs},
         year = 2020,
        month = jul,
       volume = {249},
       number = {1},
          eid = {17},
        pages = {17},
          doi = {10.3847/1538-4365/ab99c5},
archivePrefix = {arXiv},
       eprint = {1910.10395},
 primaryClass = {astro-ph.GA},
       adsurl = {https://ui.adsabs.harvard.edu/abs/2020ApJS..249...17R}
}

@ARTICLE{2014ApJ...783..105R,
       author = {{Ruan}, John J. and {Anderson}, Scott F. and {Dexter}, Jason and {Agol}, Eric},
        title = "{Evidence for Large Temperature Fluctuations in Quasar Accretion Disks from Spectral Variability}",
      journal = {\apj},
         year = 2014,
        month = mar,
       volume = {783},
       number = {2},
          eid = {105},
        pages = {105},
          doi = {10.1088/0004-637X/783/2/105},
archivePrefix = {arXiv},
       eprint = {1401.1211},
 primaryClass = {astro-ph.HE},
       adsurl = {https://ui.adsabs.harvard.edu/abs/2014ApJ...783..105R}
}

@ARTICLE{2016ApJ...826..188R,
       author = {{Ruan}, John J. and {Anderson}, Scott F. and {Cales}, Sabrina L. and {Eracleous}, Michael and {Green}, Paul J. and {Morganson}, Eric and {Runnoe}, Jessie C. and {Shen}, Yue and {Wilkinson}, Tessa D. and {Blanton}, Michael R. and {Dwelly}, Tom and {Georgakakis}, Antonis and {Greene}, Jenny E. and {LaMassa}, Stephanie M. and {Merloni}, Andrea and {Schneider}, Donald P.},
        title = "{Toward an Understanding of Changing-look Quasars: An Archival Spectroscopic Search in SDSS}",
      journal = {\apj},
         year = 2016,
        month = aug,
       volume = {826},
       number = {2},
          eid = {188},
        pages = {188},
          doi = {10.3847/0004-637X/826/2/188},
archivePrefix = {arXiv},
       eprint = {1509.03634},
 primaryClass = {astro-ph.HE},
       adsurl = {https://ui.adsabs.harvard.edu/abs/2016ApJ...826..188R}
}

@ARTICLE{2018ApJ...864...87S,
       author = {{S{\'a}nchez-S{\'a}ez}, P. and {Lira}, P. and {Mej{\'\i}a-Restrepo}, J. and {Ho}, L.~C. and {Ar{\'e}valo}, P. and {Kim}, M. and {Cartier}, R. and {Coppi}, P.},
        title = "{The QUEST-La Silla AGN Variability Survey: Connection between AGN Variability and Black Hole Physical Properties}",
      journal = {\apj},
         year = 2018,
        month = sep,
       volume = {864},
       number = {1},
          eid = {87},
        pages = {87},
          doi = {10.3847/1538-4357/aad7f9},
archivePrefix = {arXiv},
       eprint = {1808.00967},
 primaryClass = {astro-ph.GA},
       adsurl = {https://ui.adsabs.harvard.edu/abs/2018ApJ...864...87S}
}

@ARTICLE{2017ApJ...849..110S,
       author = {{S{\'a}nchez}, P. and {Lira}, P. and {Cartier}, R. and {P{\'e}rez}, V. and {Miranda}, N. and {Yovaniniz}, C. and {Ar{\'e}valo}, P. and {Milvang-Jensen}, B. and {Fynbo}, J. and {Dunlop}, J. and {Coppi}, P. and {Marchesi}, S.},
        title = "{Near-infrared Variability of Obscured and Unobscured X-Ray-selected AGNs in the COSMOS Field}",
      journal = {\apj},
         year = 2017,
        month = nov,
       volume = {849},
       number = {2},
          eid = {110},
        pages = {110},
          doi = {10.3847/1538-4357/aa9188},
archivePrefix = {arXiv},
       eprint = {1710.01306},
 primaryClass = {astro-ph.GA},
       adsurl = {https://ui.adsabs.harvard.edu/abs/2017ApJ...849..110S}
}

@article{ShakuraSunyaev1976,
  author = {Shakura, N. I. and Sunyaev, R. A.},
  title = {A theory of the instability of disk accretion on to black holes and the variability of binary X-ray sources},
  journal = {MNRAS},
  volume = {175},
  pages = {613-632},
  year = {1976}
}

@ARTICLE{1973A&A....24..337S,
       author = {{Shakura}, N.~I. and {Sunyaev}, R.~A.},
        title = "{Black holes in binary systems. Observational appearance.}",
      journal = {\aap},
         year = 1973,
        month = jan,
       volume = {24},
        pages = {337-355},
       adsurl = {https://ui.adsabs.harvard.edu/abs/1973A&A....24..337S}
}

@ARTICLE{2020ApJ...889...46S,
       author = {{Sheng}, Zhenfeng and {Wang}, Tinggui and {Jiang}, Ning and {Ding}, Jiani and {Cai}, Zheng and {Guo}, Hengxiao and {Sun}, Luming and {Dou}, Liming and {Yang}, Chenwei},
        title = "{Initial Results from a Systematic Search for Changing-look Active Galactic Nuclei Selected via Mid-infrared Variability}",
      journal = {\apj},
         year = 2020,
        month = jan,
       volume = {889},
       number = {1},
          eid = {46},
        pages = {46},
          doi = {10.3847/1538-4357/ab5af9},
archivePrefix = {arXiv},
       eprint = {1905.02904},
 primaryClass = {astro-ph.GA},
       adsurl = {https://ui.adsabs.harvard.edu/abs/2020ApJ...889...46S}
}

@ARTICLE{2017ApJ...846L...7S,
       author = {{Sheng}, Zhenfeng and {Wang}, Tinggui and {Jiang}, Ning and {Yang}, Chenwei and {Yan}, Lin and {Dou}, Liming and {Peng}, Bo},
        title = "{Mid-infrared Variability of Changing-look AGNs}",
      journal = {\apjl},
         year = 2017,
        month = sep,
       volume = {846},
       number = {1},
          eid = {L7},
        pages = {L7},
          doi = {10.3847/2041-8213/aa85de},
archivePrefix = {arXiv},
       eprint = {1707.02686},
 primaryClass = {astro-ph.GA},
       adsurl = {https://ui.adsabs.harvard.edu/abs/2017ApJ...846L...7S}
}

@ARTICLE{1985ApJ...296...46S,
       author = {{Simonetti}, J.~H. and {Cordes}, J.~M. and {Heeschen}, D.~S.},
        title = "{Flicker of extragalactic radio sources at two frequencies.}",
      journal = {\apj},
         year = 1985,
        month = sep,
       volume = {296},
        pages = {46-59},
          doi = {10.1086/163418},
       adsurl = {https://ui.adsabs.harvard.edu/abs/1985ApJ...296...46S}
}

@ARTICLE{2023ApJ...958..135S,
       author = {{Son}, Suyeon and {Kim}, Minjin and {Ho}, Luis C.},
        title = "{The Structure Function of Mid-infrared Variability in Low-redshift Active Galactic Nuclei}",
      journal = {\apj},
         year = 2023,
        month = dec,
       volume = {958},
       number = {2},
          eid = {135},
        pages = {135},
          doi = {10.3847/1538-4357/ad01bc},
archivePrefix = {arXiv},
       eprint = {2310.05389},
 primaryClass = {astro-ph.GA},
       adsurl = {https://ui.adsabs.harvard.edu/abs/2023ApJ...958..135S}
}

@INPROCEEDINGS{2005ASPC..347...29T,
   author = {{Taylor}, M.~B.},
    title = "{TOPCAT {\amp} STIL: Starlink Table/VOTable Processing Software}",
booktitle = {Astronomical Data Analysis Software and Systems XIV},
     year = 2005,
   series = {Astronomical Society of the Pacific Conference Series},
   volume = 347,
   editor = {{Shopbell}, P. and {Britton}, M. and {Ebert}, R.},
    month = dec,
    pages = {29},
   adsurl = {https://ui.adsabs.harvard.edu/abs/2005ASPC..347...29T}
}

@article{astropy,
	author = {{The Astropy Collaboration} and {Robitaille, Thomas P.} and {Tollerud, Erik J.} and {Greenfield, Perry} and {Droettboom, Michael} and {Bray, Erik} and {Aldcroft, Tom} and {Davis, Matt} and {Ginsburg, Adam} and {Price-Whelan, Adrian M.} and {Kerzendorf, Wolfgang E.} and {Conley, Alexander} and {Crighton, Neil} and {Barbary, Kyle} and {Muna, Demitri} and {Ferguson, Henry} and {Grollier, Fr\'ed\'eric} and {Parikh, Madhura M.} and {Nair, Prasanth H.} and {G\"unther, Hans M.} and {Deil, Christoph} and {Woillez, Julien} and {Conseil, Simon} and {Kramer, Roban} and {Turner, James E. H.} and {Singer, Leo} and {Fox, Ryan} and {Weaver, Benjamin A.} and {Zabalza, Victor} and {Edwards, Zachary I.} and {Azalee Bostroem, K.} and {Burke, D. J.} and {Casey, Andrew R.} and {Crawford, Steven M.} and {Dencheva, Nadia} and {Ely, Justin} and {Jenness, Tim} and {Labrie, Kathleen} and {Lim, Pey Lian} and {Pierfederici, Francesco} and {Pontzen, Andrew} and {Ptak, Andy} and {Refsdal, Brian} and {Servillat, Mathieu} and {Streicher, Ole}},
	title = {Astropy: A community Python package for astronomy},
	DOI= "10.1051/0004-6361/201322068",
	url= "https://doi.org/10.1051/0004-6361/201322068",
	journal = {A\&A},
	year = 2013,
	volume = 558,
	pages = "A33",
	month = "",
}

@ARTICLE{1997ARA&A..35..445U,
       author = {{Ulrich}, Marie-Helene and {Maraschi}, Laura and {Urry}, C. Megan},
        title = "{Variability of Active Galactic Nuclei}",
      journal = {\araa},
         year = 1997,
        month = jan,
       volume = {35},
        pages = {445-502},
          doi = {10.1146/annurev.astro.35.1.445},
       adsurl = {https://ui.adsabs.harvard.edu/abs/1997ARA&A..35..445U}
}

@ARTICLE{numpy,  
	author={S. {van der Walt} and S. C. {Colbert} and G. {Varoquaux}},  
	journal={Computing in Science   Engineering},   
	title={The NumPy Array: A Structure for Efficient Numerical Computation},   
	year={2011},  
	volume={13}, 
	number={2},  
	pages={22-30}
}

@ARTICLE{2016ApJ...829...19V,
       author = {{van Velzen}, S. and {Mendez}, A.~J. and {Krolik}, J.~H. and {Gorjian}, V.},
        title = "{Discovery of Transient Infrared Emission from Dust Heated by Stellar Tidal Disruption Flares}",
      journal = {\apj},
         year = 2016,
        month = sep,
       volume = {829},
       number = {1},
          eid = {19},
        pages = {19},
          doi = {10.3847/0004-637X/829/1/19},
archivePrefix = {arXiv},
       eprint = {1605.04304},
 primaryClass = {astro-ph.HE},
       adsurl = {https://ui.adsabs.harvard.edu/abs/2016ApJ...829...19V}
}

@ARTICLE{2004ApJ...601..692V,
       author = {{Vanden Berk}, Daniel E. and {Wilhite}, Brian C. and {Kron}, Richard G. and {Anderson}, Scott F. and {Brunner}, Robert J. and {Hall}, Patrick B. and {Ivezi{\'c}}, {\v{Z}}eljko and {Richards}, Gordon T. and {Schneider}, Donald P. and {York}, Donald G. and {Brinkmann}, Jonathan V. and {Lamb}, Don Q. and {Nichol}, Robert C. and {Schlegel}, David J.},
        title = "{The Ensemble Photometric Variability of \raisebox{-0.5ex}\textasciitilde25,000 Quasars in the Sloan Digital Sky Survey}",
      journal = {\apj},
         year = 2004,
        month = feb,
       volume = {601},
       number = {2},
        pages = {692-714},
          doi = {10.1086/380563},
archivePrefix = {arXiv},
       eprint = {astro-ph/0310336},
 primaryClass = {astro-ph},
       adsurl = {https://ui.adsabs.harvard.edu/abs/2004ApJ...601..692V}
}

@ARTICLE{2003MNRAS.345.1271V,
       author = {{Vaughan}, S. and {Edelson}, R. and {Warwick}, R.~S. and {Uttley}, P.},
        title = "{On characterizing the variability properties of X-ray light curves from active galaxies}",
      journal = {\mnras},
         year = 2003,
        month = nov,
       volume = {345},
       number = {4},
        pages = {1271-1284},
          doi = {10.1046/j.1365-2966.2003.07042.x},
archivePrefix = {arXiv},
       eprint = {astro-ph/0307420},
 primaryClass = {astro-ph},
       adsurl = {https://ui.adsabs.harvard.edu/abs/2003MNRAS.345.1271V}
}

@ARTICLE{2020NatMe..17..261V,
       author = {{Virtanen}, Pauli and {Gommers}, Ralf and {Oliphant}, Travis E. and {Haberland}, Matt and {Reddy}, Tyler and {Cournapeau}, David and {Burovski}, Evgeni and {Peterson}, Pearu and {Weckesser}, Warren and {Bright}, Jonathan and {van der Walt}, St{\'e}fan J. and {Brett}, Matthew and {Wilson}, Joshua and {Millman}, K. Jarrod and {Mayorov}, Nikolay and {Nelson}, Andrew R.~J. and {Jones}, Eric and {Kern}, Robert and {Larson}, Eric and {Carey}, C.~J. and {Polat}, {\.I}lhan and {Feng}, Yu and {Moore}, Eric W. and {VanderPlas}, Jake and {Laxalde}, Denis and {Perktold}, Josef and {Cimrman}, Robert and {Henriksen}, Ian and {Quintero}, E.~A. and {Harris}, Charles R. and {Archibald}, Anne M. and {Ribeiro}, Ant{\^o}nio H. and {Pedregosa}, Fabian and {van Mulbregt}, Paul and {SciPy 1.  0 Contributors}},
        title = "{SciPy 1.0: fundamental algorithms for scientific computing in Python}",
      journal = {Nature Medicine},
         year = 2020,
        month = feb,
       volume = {17},
        pages = {261-272},
          doi = {10.1038/s41592-019-0686-2},
archivePrefix = {arXiv},
       eprint = {1907.10121},
 primaryClass = {cs.MS},
       adsurl = {https://ui.adsabs.harvard.edu/abs/2020NatMe..17..261V}
}

@ARTICLE{2020A&A...641A.136W,
       author = {{Webb}, N.~A. and {Coriat}, M. and {Traulsen}, I. and {Ballet}, J. and {Motch}, C. and {Carrera}, F.~J. and {Koliopanos}, F. and {Authier}, J. and {de la Calle}, I. and {Ceballos}, M.~T. and {Colomo}, E. and {Chuard}, D. and {Freyberg}, M. and {Garcia}, T. and {Kolehmainen}, M. and {Lamer}, G. and {Lin}, D. and {Maggi}, P. and {Michel}, L. and {Page}, C.~G. and {Page}, M.~J. and {Perea-Calderon}, J.~V. and {Pineau}, F.-X. and {Rodriguez}, P. and {Rosen}, S.~R. and {Santos Lleo}, M. and {Saxton}, R.~D. and {Schwope}, A. and {Tom{\'a}s}, L. and {Watson}, M.~G. and {Zakardjian}, A.},
        title = "{The XMM-Newton serendipitous survey. IX. The fourth XMM-Newton serendipitous source catalogue}",
      journal = {\aap},
         year = 2020,
        month = sep,
       volume = {641},
          eid = {A136},
        pages = {A136},
          doi = {10.1051/0004-6361/201937353},
archivePrefix = {arXiv},
       eprint = {2007.02899},
 primaryClass = {astro-ph.HE},
       adsurl = {https://ui.adsabs.harvard.edu/abs/2020A&A...641A.136W}
}

@ARTICLE{2011A&A...527A..15W,
       author = {{Welsh}, B.~Y. and {Wheatley}, J.~M. and {Neil}, J.~D.},
        title = "{GALEX observations of quasar variability in the ultraviolet}",
      journal = {\aap},
         year = 2011,
        month = mar,
       volume = {527},
          eid = {A15},
        pages = {A15},
          doi = {10.1051/0004-6361/201015865},
archivePrefix = {arXiv},
       eprint = {1101.2191},
 primaryClass = {astro-ph.GA},
       adsurl = {https://ui.adsabs.harvard.edu/abs/2011A&A...527A..15W}
}

@ARTICLE{1997ApJ...475..479W,
       author = {{White}, Richard L. and {Becker}, Robert H. and {Helfand}, David J. and {Gregg}, Michael D.},
        title = "{A Catalog of 1.4 GHz Radio Sources from the FIRST Survey}",
      journal = {\apj},
         year = 1997,
        month = feb,
       volume = {475},
       number = {2},
        pages = {479-493},
          doi = {10.1086/303564},
       adsurl = {https://ui.adsabs.harvard.edu/abs/1997ApJ...475..479W}
}

@ARTICLE{2010AJ....140.1868W,
       author = {{Wright}, Edward L. and {Eisenhardt}, Peter R.~M. and {Mainzer}, Amy K. and {Ressler}, Michael E. and {Cutri}, Roc M. and {Jarrett}, Thomas and {Kirkpatrick}, J. Davy and {Padgett}, Deborah and {McMillan}, Robert S. and {Skrutskie}, Michael and {Stanford}, S.~A. and {Cohen}, Martin and {Walker}, Russell G. and {Mather}, John C. and {Leisawitz}, David and {Gautier}, III, Thomas N. and {McLean}, Ian and {Benford}, Dominic and {Lonsdale}, Carol J. and {Blain}, Andrew and {Mendez}, Bryan and {Irace}, William R. and {Duval}, Valerie and {Liu}, Fengchuan and {Royer}, Don and {Heinrichsen}, Ingolf and {Howard}, Joan and {Shannon}, Mark and {Kendall}, Martha and {Walsh}, Amy L. and {Larsen}, Mark and {Cardon}, Joel G. and {Schick}, Scott and {Schwalm}, Mark and {Abid}, Mohamed and {Fabinsky}, Beth and {Naes}, Larry and {Tsai}, Chao-Wei},
        title = "{The Wide-field Infrared Survey Explorer (WISE): Mission Description and Initial On-orbit Performance}",
      journal = {\aj},
         year = 2010,
        month = dec,
       volume = {140},
       number = {6},
        pages = {1868-1881},
          doi = {10.1088/0004-6256/140/6/1868},
archivePrefix = {arXiv},
       eprint = {1008.0031},
 primaryClass = {astro-ph.IM},
       adsurl = {https://ui.adsabs.harvard.edu/abs/2010AJ....140.1868W}
}

@ARTICLE{2018ApJ...862..109Y,
       author = {{Yang}, Qian and {Wu}, Xue-Bing and {Fan}, Xiaohui and {Jiang}, Linhua and {McGreer}, Ian and {Shangguan}, Jinyi and {Yao}, Su and {Wang}, Bingquan and {Joshi}, Ravi and {Green}, Richard and {Wang}, Feige and {Feng}, Xiaotong and {Fu}, Yuming and {Yang}, Jinyi and {Liu}, Yuanqi},
        title = "{Discovery of 21 New Changing-look AGNs in the Northern Sky}",
      journal = {\apj},
         year = 2018,
        month = aug,
       volume = {862},
       number = {2},
          eid = {109},
        pages = {109},
          doi = {10.3847/1538-4357/aaca3a},
archivePrefix = {arXiv},
       eprint = {1711.08122},
 primaryClass = {astro-ph.GA},
       adsurl = {https://ui.adsabs.harvard.edu/abs/2018ApJ...862..109Y}
}

@ARTICLE{2025ApJ...993..203T,
       author = {{Tomar}, Ashutosh and {Rakshit}, Suvendu and {Mandal}, Amit Kumar and {Pandey}, Shivangi},
        title = "{Probing the Dust Torus Radius─Luminosity Relation: A WISE View}",
      journal = {\apj},
         year = 2025,
        month = nov,
       volume = {993},
       number = {2},
          eid = {203},
        pages = {203},
          doi = {10.3847/1538-4357/ae0f96},
archivePrefix = {arXiv},
       eprint = {2510.04517},
 primaryClass = {astro-ph.GA},
       adsurl = {https://ui.adsabs.harvard.edu/abs/2025ApJ...993..203T}
}

@ARTICLE{2024ApJ...968...59M,
       author = {{Mandal}, Amit Kumar and {Woo}, Jong-Hak and {Wang}, Shu and {Rakshit}, Suvendu and {Cho}, Hojin and {Son}, Donghoon and {Stalin}, C.~S.},
        title = "{Revisiting the Dust Torus Size─Luminosity Relation Based on a Uniform Reverberation-mapping Analysis}",
      journal = {\apj},
         year = 2024,
        month = jun,
       volume = {968},
       number = {2},
          eid = {59},
        pages = {59},
          doi = {10.3847/1538-4357/ad414d},
archivePrefix = {arXiv},
       eprint = {2403.01885},
 primaryClass = {astro-ph.GA},
       adsurl = {https://ui.adsabs.harvard.edu/abs/2024ApJ...968...59M}
}

@ARTICLE{2020ApJ...900...58Y,
       author = {{Yang}, Qian and {Shen}, Yue and {Liu}, Xin and {Aguena}, Michel and {Annis}, James and {Avila}, Santiago and {Banerji}, Manda and {Bertin}, Emmanuel and {Brooks}, David and {Burke}, David and {Carnero Rosell}, Aurelio and {Carrasco Kind}, Matias and {da Costa}, Luiz and {De Vicente}, Juan and {Desai}, Shantanu and {Diehl}, H. Thomas and {Doel}, Peter and {Flaugher}, Brenna and {Fosalba}, Pablo and {Frieman}, Josh and {Garcia-Bellido}, Juan and {Gerdes}, David and {Gruen}, Daniel and {Gruendl}, Robert and {Gschwend}, Julia and {Gutierrez}, Gaston and {Hinton}, Samuel and {Hollowood}, Devon L. and {Honscheid}, Klaus and {Kuropatkin}, Nikolay and {Maia}, Marcio and {March}, Marisa and {Marshall}, Jennifer and {Martini}, Paul and {Melchior}, Peter and {Menanteau}, Felipe and {Miquel}, Ramon and {Paz-Chinchon}, Francisco and {Malag{\'o}n}, Andr{\'e}s Plazas and {Romer}, Kathy and {Sanchez}, Eusebio and {Scarpine}, Vic and {Schubnell}, Michael and {Serrano}, Santiago and {Sevilla}, Ignacio and {Smith}, Mathew and {Suchyta}, Eric and {Tarle}, Gregory and {Varga}, Tamas Norbert and {Wilkinson}, Reese},
        title = "{Dust Reverberation Mapping in Distant Quasars from Optical and Mid-infrared Imaging Surveys}",
      journal = {\apj},
         year = 2020,
        month = sep,
       volume = {900},
       number = {1},
          eid = {58},
        pages = {58},
          doi = {10.3847/1538-4357/aba59b},
archivePrefix = {arXiv},
       eprint = {2007.02402},
 primaryClass = {astro-ph.GA},
       adsurl = {https://ui.adsabs.harvard.edu/abs/2020ApJ...900...58Y}
}

@ARTICLE{2023ApJ...950..122L,
       author = {{Li}, Junyao and {Shen}, Yue},
        title = "{Constraining AGN Torus Sizes with Optical and Mid-infrared Ensemble Structure Functions}",
      journal = {\apj},
         year = 2023,
        month = jun,
       volume = {950},
       number = {2},
          eid = {122},
        pages = {122},
          doi = {10.3847/1538-4357/accade},
archivePrefix = {arXiv},
       eprint = {2302.12437},
 primaryClass = {astro-ph.GA},
       adsurl = {https://ui.adsabs.harvard.edu/abs/2023ApJ...950..122L}
}

@ARTICLE{2026ApJS..285...20P,
       author = {{Paliya}, Vaidehi S. and {Rakshit}, Suvendu and {Dom{\'\i}nguez}, Alberto and {Saikia}, D.~J. and {Stalin}, C.~S.},
        title = "{Narrow-line Seyfert 1 Galaxies in the Dark Energy Spectroscopic Instrument Data Release 1}",
      journal = {\apjs},
         year = 2026,
        month = jul,
       volume = {285},
       number = {1},
          eid = {20},
        pages = {20},
          doi = {10.3847/1538-4365/ae7c6f},
archivePrefix = {arXiv},
       eprint = {2606.10447},
 primaryClass = {astro-ph.GA},
       adsurl = {https://ui.adsabs.harvard.edu/abs/2026ApJS..285...20P}
}

@misc{camus2026compilinglargestmultiwavelengthdatabase,
      title={Compiling the largest multiwavelength database of known changing-look AGN: Prospects in the Rubin era}, 
      author={Mariangella Camus and Swayamtrupta Panda},
      year={2026},
      eprint={2607.25839},
      archivePrefix={arXiv},
      primaryClass={astro-ph.GA},
      url={https://arxiv.org/abs/2607.25839}, 
}

@ARTICLE{2026ApJ...999..243F,
       author = {{Franz}, Noah and {Alexander}, Kate D. and {Gomez}, Sebastian and {Christy}, Collin T. and {Laskar}, Tanmoy and {van Velzen}, Sjoert and {Earl}, Nicholas and {Gezari}, Suvi and {Karmen}, Mitchell and {Margutti}, Raffaella and {Pearson}, Jeniveve and {Villar}, V. Ashley and {Zabludoff}, Ann I.},
        title = "{The Open mulTiwavelength Transient Event Repository (OTTER): Infrastructure Release and Tidal Disruption Event Catalog}",
      journal = {\apj},
         year = 2026,
        month = mar,
       volume = {999},
       number = {2},
          eid = {243},
        pages = {243},
          doi = {10.3847/1538-4357/ae346e},
archivePrefix = {arXiv},
       eprint = {2509.05405},
 primaryClass = {astro-ph.HE},
       adsurl = {https://ui.adsabs.harvard.edu/abs/2026ApJ...999..243F}
}
\bibliographystyle{aasjournalv7}

\end{document}